\documentclass[
reprint,
 amsmath,amssymb,
 aps,
 pre,
 superscriptaddress,
 longbibliography,
]{revtex4-2}

\usepackage{xcolor}
\usepackage{enumitem}
\usepackage{graphicx}
\usepackage{dcolumn}
\usepackage{verbatim}
\usepackage{bm}

\usepackage[margin=1.5cm]{geometry}

\usepackage[scr=boondox,  
            cal=esstix]   
           {mathalpha}

\usepackage{amsmath,amsfonts,amsthm}
\usepackage{enumitem}
\usepackage[caption=false, subrefformat=parens,labelformat=parens]{subfig}
\newtheorem{theorem}{Theorem}

\newtheorem{lemma}[theorem]{Lemma}
\usepackage{enumitem}
\usepackage[normalem]{ulem}

\newcommand{\p}{\mathbb{P}}

\newcommand{\E}{\mathbb{E}}

\newcommand{\T}{\mathsf{T}}
\newcommand{\1}{\mathbf{1}}
\newcommand{\A}{\mathcal{A}}
\newcommand{\B}{\mathcal{B}}
\newcommand{\subG}{\mathsf{subG}}
\newcommand{\R}{\sigma_r}

\begin{document}

\title{Resetting-based designated information distribution}

\author{Xiaojuan Ma}
\thanks{These authors contributed equally to this work and are listed alphabetically.}
\affiliation{
Siebel School of Computing and Data Science, University of Illinois Urbana-Champaign, Champaign, Illinois 61820, USA}

\author{Yan Wang}
\thanks{These authors contributed equally to this work and are listed alphabetically.}
\affiliation{
Department of Mathematics, Wayne State University, Detroit, Michigan 48202, USA }
\affiliation{Institute for AI and Data Science (AIDaS), Wayne State University, Detroit, Michigan 48202, USA}

\author{Indranil Gupta}
\affiliation{
Siebel School of Computing and Data Science, University of Illinois Urbana-Champaign, Champaign, Illinois 61820, USA}

\begin{abstract}  
    For a network of $n$ nodes under random gossip dynamics, we show that by introducing the additional resetting mechanism (with rate $\beta$) to a designated subset of $r$ nodes, the expected sum of all node values is asymptotically equal to $\tfrac{n}{r}$ times the sum of the initial values of the $r$ designated nodes, in the small-$\beta$ and long-time limit. We also analyze the convergence speed to this asymptotic state and provide an approximate high-probability confidence interval for the expected sum at finite time $t$ and finite $\beta$. Operationally, we propose a protocol that distributes the initial node values over the designated subset evenly (up to a prescribed precision),
    by appropriately turning on and off the resetting dynamics. Our work not only offers insights into classical problems such as distributed computing and network control, but also serves as a potential foundation for designing quantum algorithms that prepare desired states in quantum networks using only local operations.  

\end{abstract}

\maketitle

\section{Introduction}
\label{sec:introduction}

Consider a connected network with $n$ nodes. At each time $t\geq 0$, the network carries some information, which is characterized by a state vector $x_t$ with $x_t(i)$ being the state value of the $i$th node. This state vector can, for example, describe the spin configuration of a physical system, where $x_t(i) \in \{\pm 1\}$ \cite{kulske2025ising, mobilia2003does}; the voting pattern in a consensus model, where $x_t(i)$ may take values in one of possible states or even be a general probability distribution \cite{clifford1973model, holley1975ergodic, castellano2009statistical, starnini2012ordering, bohme2012fragmentation, degroot1974reaching}; the sensor readings in a sensor network \cite{estrin1999next, heinzelman1999adaptive, sengul2008adaptive}; or a statistic of model weights in decentralized federated learning, where $x_t(i)\in \mathbb{R}$ \cite{lalitha2018fully, lian2017can, hegedHus2019gossip, yuan2024decentralized}. In such settings, adjacent nodes typically exchange their individual information over time, and this process can be effectively captured by the gossip dynamics \cite{boyd2006randomized, boyd2005gossip, kempe2004spatial, haas2002gossip, jelasity2005gossip} where each node updates its value at time $t$, $x_t(i)$, based on interactions with its neighbors. 
{Under fairly general update rules of the gossip dynamics, conformity among components will emerge, giving rise to the ferromagnetic phase in spin systems, consensus in social systems, and synchronization in dynamical systems \cite{baxter2016exactly, newman2018networks, olfati2007consensus, strogatz2000kuramoto, barabasi2013network}.} In the setting where $x_t(i)\in\mathbb{R}$, it is well established that the network will eventually achieve the synchronized state \cite{boyd2006randomized}, which is given by the average of the initial state of all nodes:
\begin{align}\label{eq:gossip_do}
x_\infty(1)=\cdots=x_\infty(n)=\frac{1}{n}\sum_{i=1}^n x_0(i) \quad \text{(by gossip).}
\end{align}
This result of average consensus also has important implications for distributed computing. It shows that global information sharing can emerge from purely local dynamics \cite{shah2009gossip, kempe2003gossip, boyd2006randomized}. In particular, average consensus is sufficient for evaluating any
linear function that does not depend on node labels. If
\(F:\mathbb{R}^n\to\mathbb{R}\) is linear and invariant under
permutations of the node values, then there exists a coefficient
vector \(a=(a_1,\ldots,a_n)^\top\in\mathbb{R}^n\) such that
\(F(x)=a^\top x\) for all \(x\in\mathbb{R}^n\), and swapping any two coordinates \(i\) and \(j\)
gives
\(
(a_i-a_j)(x_i-x_j)=0
\)
for all \(x\). Therefore, \(a_i=a_j\) for all \(i,j\), and
\(F(x)\propto \sum_{i=1}^n x(i)\) is a scalar multiple of the sum. Hence, there exists a scalar
function \(f:\mathbb{R}\to\mathbb{R}\) such that
\(
F(x)=f\!\left(\frac{1}{n}\sum_{i=1}^n x(i)\right).
\)
Thus, once nodes agree on the mean, no further internode
communication is needed to evaluate such an \(F\)
\cite{mazzarella2015consensus}.

Average consensus and its variants have been widely studied; see, for example,~\cite{olfati2004consensus, olfati2007consensus, boyd2006randomized}. These methods naturally arise in fully decentralized settings, where no centralized or privileged nodes exist and communication is restricted to local interactions. They underpin applications such as clock synchronization~\cite{carli2008pi, schenato2011average}, decentralized learning~\cite{lian2017can}, and distributed optimization~\cite{boyd2011distributed,nedic2009distributed}.

A distinctive feature of the consensus state, as shown in
Eq.~(\ref{eq:gossip_do}), is that every node in the network
contributes equally to the final average. In practice, however,
many distributed computing tasks concern only a designated subset
of node values. In such settings, all nodes may still participate
in the communication process, while the computed result should
depend only on the initial values of the designated subset. This
distinction between communication and contribution arises naturally
in distributed information processing. For example, in sensor networks, one may wish to average readings
only from sensors satisfying a prescribed condition, such as sensors
in a certain region or sensors detecting a particular event, while
other nodes merely relay the information through the network
\cite{madden2005tinydb}. Another example is
federated learning, where, in each communication round, model updates from a selected
subset of participating clients are aggregated~\cite{mcmahan2017communication}.

This naturally raises the question of whether information within a designated subset of nodes, embedded in a larger network, can be shared solely through local operations. Note that the nodes in this designated subset are not necessarily directly connected and form a cluster, and one is not allowed 
to create new shortcuts between them or to remove existing links in the network. In such cases, the gossip dynamics alone is not enough (though it remains essential for disseminating information across the network and should be kept as one ingredient of the protocol), and another mechanism is needed to ``amplify'' the signal originating from the designated subset of nodes. We show that this role is played by a local \emph{resetting} operation applied only to the designated nodes.

Our protocol relies on a crucial observation from Eq. (\ref{eq:gossip_do}): if the goal is to distribute some quantity uniformly across the network (computing the average), it is the \emph{sum} of the initial values $x_0(i)$, rather than the initial state vector $x_0$ itself, that really matters under the gossip dynamics. In this work, we consider the case where an ambient network undergoes the gossip dynamics, and we only have access to a given subset of nodes, i.e., we can perform local operations on them. We do not make any assumptions about the underlying topology or the locations of the designated nodes in the network.

Without loss of generality, we let such nodes be indexed by $\{1,\ldots,r\}$, where $1\leq r \leq n$ is the size of the subset. Recall that our goal is to drive the designated subset of nodes to reach a state primarily determined by their own initial states, with only a secondary (ideally vanishing) influence from the rest of the network. That is,
\begin{align}\label{eq:goal}
x_\infty(1)=\cdots=x_\infty(r)\approx \frac{1}{r}\sum_{i=1}^r x_0(i)\quad \text{(our goal).}
\end{align}
Based on the observation above, our goal (\ref{eq:goal}) can essentially be achieved by requiring that
\begin{align}\label{eq:goal_real}
    \sum_{i=1}^n x_{t}(i)\approx \frac{n}{r}\sum_{i=1}^r x_0(i).
\end{align}
In other words, the problem can be formulated as controlling the component sum $\sum_{i=1}^n x_t(i)$ via appropriate operations. Once the sum is brought sufficiently close to the desired value specified in Eq. (\ref{eq:goal_real}), the operation is then discontinued. The whole network will subsequently evolve under the gossip dynamics alone to a state that satisfies condition (\ref{eq:goal}), as is well known.

In this work, $x_0(i)$'s are fixed quantities, but all $x_t(i)$'s for $t>0$ are random as a result of our protocol. To achieve (\ref{eq:goal_real}), we require that at some time $t$ specified by the protocol, the sum $\sum_{i=1}^n x_t(i)$ is sufficiently close to $\frac{n}{r}\sum_{i=1}^r x_0(i)$ with high probability. The precise meaning of (\ref{eq:goal_real}) is given in Sec.~\ref{subsec:probalisitc_bound_for_single_trail}.

One major contribution of our work is that we uncover resetting as the crucial mechanism to achieve Eq. (\ref{eq:goal_real}).
Resetting has been shown to lead to nontrivial dynamical consequences  in various situations, including random search \cite{pal2017first, singh2024competing, biswas2024search}, Bernoulli trials \cite{
belan2018restart,singh2023bernoulli}, first-passage problems in diffusion processes \cite{de2020optimization}, 
and generative machine learning \cite{xu2023restart}. In the study of opinion dynamics, this mechanism has been used to model a zealot in the voter model \cite{mobilia2003does}, where
the presence of a single node with the resetting mechanism (a special case of our work) suffices to drive the whole system into the consensus state determined by its own initial state. In the Friedkin-Johnsen (FJ) model \cite{friedkin1990social,ghaderi2014opinion, xing2024concentration,wang2024final}, each node may persistently hold its initial opinion to some extent. The resetting mechanism modifies the long-time limit of the system state, resulting in a final state in which each node converges to a weighted sum of the stubborn nodes' initial values, with the weights determined by factors such as the network structure and the placement and stubbornness of the stubborn nodes.

\emph{Intuition of the protocol.} 
Our protocol can be represented in part by a specific randomized FJ model. However, there are three fundamental differences between our work and prior work on the FJ model. First, our ultimate goal is to find a local operation (which turns out to be resetting) such that the system can be approximately driven to a desired state, rather than to investigate the outcome of the \emph{given} resetting operation on the opinion dynamics. Second, as will be shown below, the relevant region of the {resetting rate $\beta$ (defined in Sec. \ref{sec:model} {as the probability that a designated node is reset to its initial value at each time step}) } in our work is $\beta\ll1$. Even when $\beta$ is vanishingly small, the effect of the seemingly weak resetting dynamics still has a profound impact on the long-time behavior of the system, based on which Eq. (\ref{eq:goal_real}) is established. Third, our protocol includes operational guidance as to the intensity and duration of the resetting dynamics, which is beyond the scope of the FJ model.   

In the following sections, we provide a self-contained analysis of the resetting-modified gossip dynamics. 
\begin{enumerate}[label=(\roman*)]
    \item We first derive the long-time limit of the mean sum of the state vector components $\E[\sum_{i=1}^nx_\infty(i)]$, which is well approximated by $\frac{n}{r}\sum_{i=1}^r x_0(i)$, and the difference is proportional to $\beta$. Physically, this result shows that, for small $\beta$, the whole network will eventually be dominated by the designated subset of nodes.
    \item We also show that the convergence of the finite-time mean sum $\E[\sum_{i=1}^n x_t(i)]$ to its limit is exponentially fast and obtain an approximate rate in terms of $\beta$.
    \item Finally, conditional on a given network, we establish an approximate high-probability bound for a random realization of the process, where with probability at least $1-\delta$ [$\delta\in(0,1)$ is prescribed], the (random) sum $\sum_{i=1}^n x_t(i)$ at or beyond some critical time $t_\ast$ is close to the desired value of $\frac{n}{r}\sum_{i=1}^r x_0(i)$ within a prescribed precision $\varepsilon$, as long as we set $\beta\leq\beta^\ast$, where $\beta^\ast$ is a critical value of the resetting rate. Both $t_\ast$ and $\beta^\ast$ can be calculated from our theory.
\end{enumerate}

Based on these results, our protocol proceeds in two steps. First, we generate the desired sum of $\sum_{i=1}^nx_t(i)$ by running the resetting dynamics (together with the gossip dynamics) at a proper rate for an appropriate duration. Then, we turn off the resetting dynamics, and the sum is distributed over the whole network by the gossip dynamics alone. We stress that controlling the sum 
$\sum_{i=1}^n x_t(i)$ to the vicinity of the target, rather than a specific state value distribution over the network, is sufficient for the two-step protocol to achieve uniform value distribution across all nodes. This is because the gossip dynamics in the second step guarantees that every node 
converges to the same value, regardless of how values are distributed across nodes at the start of the second step. Since the gossip dynamics has been extensively studied and well understood, we only focus on the first step of our protocol to achieve Eq. (\ref{eq:goal_real}).

From a broader perspective, our results are also pertinent to quantum settings where it is desirable to prepare a common quantum state among subsystems in a controllable and distributed way. Due to the no-cloning theorem and the destructive nature of measurements, admissible protocols should avoid information replication or value inspection. The classical counterpart investigated in this work may thus serve as a foundation for developing a fully quantum protocol. We leave this for future work.

We present the details of our dynamical model in Sec.~\ref{sec:model}. Theoretical results and numerical verifications are given in Secs.~\ref{sec:results} and \ref{sec:experiments}, respectively. We discuss applications of our protocol in more general settings in Sec.~\ref{sec:discussion}, and conclude our work in Sec.~\ref{sec:conclusion}.

\section{Dynamical model}\label{sec:model}
\begin{figure}
    \centering
    \includegraphics[width=.85\linewidth]{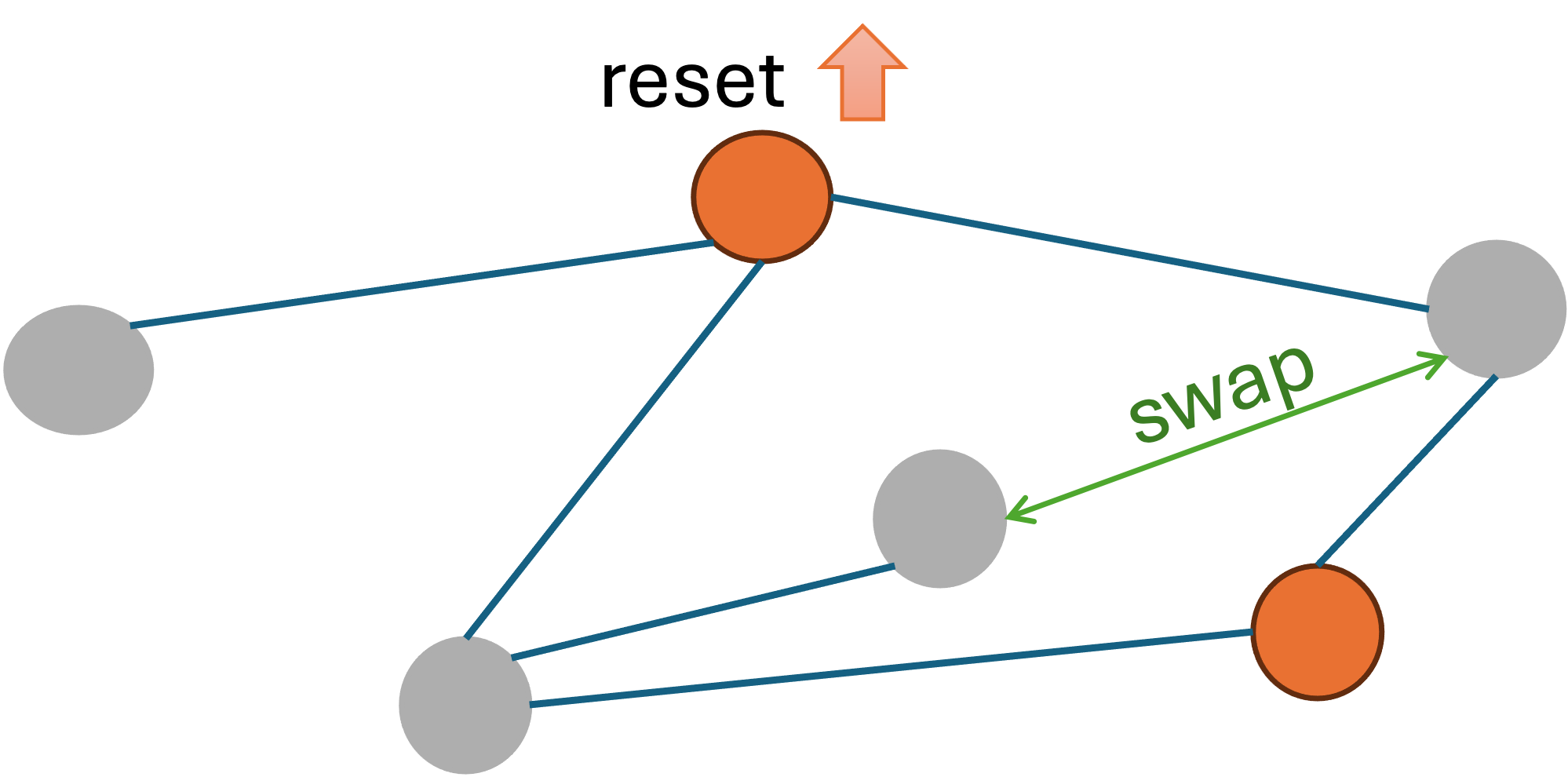}
    \caption{Model illustration: In an underlying connected network with arbitrary topology, a randomly selected pair of connected nodes probabilistically swap their states at each time step (with rate $\alpha$). Subsequently, one node from the designated set (highlighted in orange) probabilistically resets its state to its initial value (with rate $\beta$). Note that the designated nodes may be arbitrarily distributed across the network and do not necessarily form a cluster. 
    }
    \label{fig:setting_illustration}
\end{figure}

Suppose we are given a connected network with $n$ nodes. The set of edges consists of unordered pairs $(i,j)$, which means there is a direct link between nodes $i$ and $j$:
\[
E=\{(i,j)\}.
\]
The cardinality of $E$ is denoted as $m$. The adjacency matrix of the network is a symmetric $n\times n$ matrix $A$ with $A_{ij}=1$ if $(i,j)\in E$ and $A_{ij}=0$ if $(i,j)\notin E$. For all $i\in[n]$, we have $A_{ii}=0$. (We use $[n]$ to denote the set $\{1,\ldots,n\}$.) At time $t=0,1,2,\ldots$, a vector $x_t$ of length $n$ is used to represent the state of the network, and the state of the $i$th node is $x_t(i)$. Without loss of generality, we let $[r]$ with $1\leq r\leq n$ be the index set of designated nodes to which the additional resetting dynamics is applied. The set $[n]\backslash [r]$ is the index set of $s=n-r$ remaining nodes in the network that are not subject to resetting. Aside from requiring the graph to be connected, we make no assumptions about its topology or the locations of the designated nodes.

As illustrated in Fig.~\ref{fig:setting_illustration}, there are two ingredients of the dynamical process in our model: swapping and resetting.

(1) \emph{Swapping}. This gives rise to the gossip dynamics. At time $t$, a uniformly randomly selected pair of nodes will exchange their states. That is, some edge $(i_t,j_t)\in E$ is uniformly randomly picked such that nodes $i_t$ and $j_t$ will (partially) exchange their information by swapping their states
\[
    (x_t(j_t),x_t(i_t)) \leftarrow (x_t(i_t),x_t(j_t)).
\]
The intensity of this swapping process is denoted by some $\alpha\in(0,1)$. Physically, this can be interpreted as the coupling intensity between nodes $i_t$ and $j_t$; probabilistically, one may understand that swapping happens with probability $\alpha$, and the system is unchanged with probability $1-\alpha$. In the matrix form, this process is given by
\begin{align}\label{eq:swap}
    \tilde x_t = (1-\alpha)x_t + \alpha W_tx_t && \text{(swapping)}, 
\end{align}
where $\tilde x_t$ denotes the intermediate state after the swapping process and 
\[
    W_t = I_n - (e_{i_t}-e_{j_t})(e_{i_t}-e_{j_t})^\T
\]
is a random matrix with $(i_t,j_t)\in E$ being the edge selected at time $t$, $I_n$ is the $n\times n$ identity matrix, $e_{i}$ is $i$th standard orthonomral basis vector in the $n$-dimensional Euclidean space $\mathbb{R}^n$, and the operation of transpose is denoted by the superscript $^\T$. 

The swapping process serves as the core of classical gossip dynamics and its properties are well known \cite{boyd2006randomized}. First, we know that
\begin{align}\label{eq:EW}
    \E[W_t] = I_n-\frac{L}{m},
\end{align}
where $L=D-A$ is the Laplacian matrix of the network and $D=\text{diag}\{d_1,\ldots,d_n\}$ is the degree matrix with $d_i=\sum_{j=1}^n A_{ij}$. The expectation is with respect to the random edge selection process. Second, swapping does not change the sum of the state vector components. That is, for all $t$, 
\[
    \1_n^\T W_t x_t = \1_n^\T x_t,
\]
where $\1_n$ is the $n$-dimensional vector with all entries being 1. 

(2) \emph{Resetting}. After the swapping process, an independent resetting process is triggered. The state of a uniformly randomly selected node indexed in $[r]$ will be reset to its initial state at rate $\beta$. That is, for the selected node $k_t\in[r]$, we update its state as
\begin{align}\label{eq:resetting}
x_{t+1}(k_t) = (1-\beta) \tilde x_t(k_t) + \beta x_0(k_t)  && \text{(resetting)}. 
\end{align}
The remaining nodes keep their states unchanged. 

We stress that $k_t$ is not necessarily equal to $i_t$ or $j_t$ in the swapping process. The edge and node selections are independent. With the help of an ancillary state vector that always stays at $x_0$, we rewrite the resetting dynamics as 

\begin{widetext}
\begin{align}\label{eq:reset}
    \begin{bmatrix}
        x_{t+1} \\
        x_0
    \end{bmatrix} 
    & = 
    \begin{bmatrix}
        (1-\beta)I_r+\beta(I_r- e_{k_t|r}e_{k_t|r}^\T) & 0_{r\times s} & \beta e_{k_t|r}e_{k_t|r}^\T & 0_{r\times s}\\
        0_{s\times r} & I_s & 0_{s\times r} & 0_{s\times s}\\
        0_{r\times r} & 0_{r\times s} & I_r & 0_{r\times s}\\
        0_{s\times r} & 0_{s\times s} & 0_{s\times r} & I_s
    \end{bmatrix}
    \begin{bmatrix}
        \tilde x_t\\
        x_0
    \end{bmatrix}
    := P_t\begin{bmatrix}
        \tilde x_t\\
        x_0
    \end{bmatrix},
\end{align}
\end{widetext}
where $e_{k|r}$ denotes the $k$th standard orthonormal basis vector in the $r$-dimensional Euclidean space $\mathbb{R}^r$ and $0_{s\times r}$ is an $s\times r$ matrix with all entries being 0.

Combining the two processes (\ref{eq:swap}) and (\ref{eq:reset}) together yields the complete state update rule at time $t$:
\begin{align}\label{eq:state_update}
    \begin{bmatrix}
        x_{t+1} \\
        x_0
    \end{bmatrix} 
    &= 
    P_t
    \begin{bmatrix}
        (1-\alpha)I_n + \alpha W_t & 0_{n\times n}\\
        0_{n\times n} & I_n
    \end{bmatrix}
    \begin{bmatrix}
        x_t\\
        x_0
    \end{bmatrix}\notag\\
    &:= P_tQ_t\begin{bmatrix}
        x_t\\
        x_0
    \end{bmatrix}.
\end{align}
Applying Eq. (\ref{eq:state_update}) iteratively yields
\begin{align*}
    \begin{bmatrix}
        x_{t+1} \\
        x_0
    \end{bmatrix} & = P_tQ_t\ldots P_0Q_0
    \begin{bmatrix}
        x_{0} \\
        x_0
    \end{bmatrix}:=\left(\prod_{\tau=0}^t P_\tau Q_\tau\right)
    \begin{bmatrix}
        x_{0} \\
        x_0
    \end{bmatrix}.
\end{align*}
Note that $P_\tau$ is a block upper triangular matrix and $Q_\tau$ is a block-diagonal matrix. The product $P_\tau Q_\tau$ thus is also a block upper triangular matrix and we denote it as
\[
P_\tau Q_\tau = \begin{bmatrix}
    \A_\tau & \B_\tau\\
    0_{n\times n} & I_n
\end{bmatrix},
\]
where $\A_\tau = (I_n-\beta e_{k_\tau}e_{k_\tau}^\T)[I_n-\alpha(e_{i_\tau}-e_{j_\tau})(e_{i_\tau}-e_{j_\tau})^\T]$, $\B_\tau=\beta e_{k_\tau}e_{k_\tau}^\T$, and $e_i$ is the $i$th standard basis vector in the Euclidean space $\mathbb{R}^n$. Therefore,
\begin{align*}
    \left( \prod_{\tau=0}^{t-1} P_\tau Q_\tau \right) =  \begin{bmatrix}
    \mathbb{A}_{t-1} & \mathbb{B}_{t-1}\\
    0_{n\times n} & I_n
\end{bmatrix},
\end{align*}
where $\mathbb{A}_{t-1} = \A_{t-1}\ldots \A_0$ and
\begin{align*}
\mathbb{B}_{t-1} &= \A_{t-1}\ldots \A_{1}\B_0 + \ldots + \A_{t-1}\B_{t-2}+\B_{t-1}.
\end{align*}
We finally have an expression for $x_t$ as
\begin{align}\label{eq:xt}
    x_t
    &= (\mathbb{A}_{t-1}+\mathbb{B}_{t-1})x_0\notag\\
    &= (\A_{t-1}\ldots\A_0 + \A_{t-1}\ldots\A_1\B_0+\cdots+\B_{t-1})x_0.
\end{align}
This equation serves as the foundation for our analysis, as detailed below. We treat $x_0$ as fixed, and $x_t$ is a random vector due to the randomness in the matrix sequences $\{\A_\tau\}$ and $\{\B_\tau\}$.

\section{Results}\label{sec:results}
In this section, we first derive the asymptotic mean of the sum of state values under the swapping and resetting dynamics in Sec.~\ref{subsec:asymptotic_mean_state}. We then establish an approximate convergence rate of the finite-time mean sum to its asymptotic value in Sec.~\ref{subsec:finit_time_convergence_rate}. Finally, in Sec.~\ref{subsec:probalisitc_bound_for_single_trail}, based on an effective dynamics, we investigate how the finite-time sum of state values from a random realization deviates from the asymptotic mean, and establish a high-probability bound for this deviation, which provides guidance on the choice of key parameters for our protocol.

\subsection{Asymptotic mean state}
\label{subsec:asymptotic_mean_state}
Starting from a given initial state vector $x_0$ and network Laplacian $L$, each realization of the process leads to a random final state vector $x_t$. We first analyze the mean of $x_t$, denoted as $\E[x_t]$, where the expectation is taken over all the randomness in the process but conditional on $x_0$ and $L$. To this end, we will derive a closed-form expression for $\E[x_t]$.

Since the swapping and resetting dynamics are independent and identically distributed at each round, we might omit the subscript in $W_\tau,\A_\tau,\B_\tau,P_\tau$, and $Q_\tau$ when calculating their means. By independence, we have
    $\E[PQ] = \E[P]\E[Q]$. By the definition of $Q$, 
\[
\E[Q] = \begin{bmatrix}
    (1-\alpha)I_n + \alpha \E[W] & 0_{n\times n}\\
    0_{n\times n} & I_n
\end{bmatrix}.
\]
Inserting Eq. (\ref{eq:EW}) to the expression above yields
\[
(1-\alpha)I_n + \alpha \E[W] = I_n - \frac{\alpha}{m}L.
\]
Hence we conclude
\[
\E[Q] = \begin{bmatrix}
        I_n -\frac{\alpha}{m}L & 0_{n\times n}\\
        0_{n\times n} & I_n
    \end{bmatrix}.
\]
Similarly, by the definition of $P$ in Eq. (\ref{eq:reset}), one of the $r$ nodes is selected uniformly random at each round. That is, $k_t$ is uniformly randomly distributed in $[r]$, i.e., $k_t\sim\mathsf{Unif}([r])$. Taking expectation with respect to $k$ yields
\[
\E_{k_t\sim\mathsf{Unif}([r])}[e_{k_t|r}e_{k_t|r}^\T] = \frac{1}{r}\left(\sum_{k=1}^r e_{k|r}e_{k|r}^\T\right) = \frac{1}{r}I_r,
\]
and we thus have  
\begin{align*}
    \E[P] & = \begin{bmatrix}
        \left(1-\frac{\beta}{r}\right)I_r & 0_{r\times s} & \frac{\beta}{r} I_r & 0_{r\times s}\\
        0_{s\times r} & I_s & 0_{s\times r} & 0_{s\times s}\\
        0_{r\times r} & 0_{r\times s} & I_r & 0_{r\times s}\\
        0_{s\times r} & 0_{s\times s} & 0_{s\times r} & I_s
    \end{bmatrix}\\
    &:= \begin{bmatrix}
       I_n - \frac{\beta}{r} UU^\T &  \frac{\beta}{r} UU^\T \\
       0_{n\times n} & I_n
    \end{bmatrix},
\end{align*}
where $U=[e_1,\ldots,e_r]$ is an $n\times r$ matrix.

The explicit expressions for $\E[\A]$ and $\E[\B]$ can be obtained based on $\E[P]$ and $\E[Q]$ as
\begin{align}\label{eq:EA}
\E[\A] = 
    \underbrace{\left(I_n - \frac{\beta}{r}UU^\T\right)}_{:=T_1}
\underbrace{\left(
I_n - \frac{\alpha}{m}L
\right)}_{:=T_2},
\end{align}
and 
\begin{align}\label{eq:EB}
\E[\B] = \frac{\beta}{r}UU^\T.
\end{align}
Again, by the independence of the dynamics at different times, we have
\[
\E[\mathbb{A}_{t-1}]= \E[\A]^t
\]
and
\[
\E[\mathbb{B}_{t-1}] = (\E[\A]^{t-1}+\cdots +I_n)\E[\B].
\]
Taking the expectation to Eq. (\ref{eq:xt}) gives the mean state vector at time $t$ as
\begin{align}\label{eq:Ext} 
    \E[x_t] &= \E[\A]^t x_0 + (\E[\A]^{t-1}+\cdots+ I_n)\E[\B] x_0.
\end{align}
We are interested in the $t\to\infty$ limit of \[y_t:=\1_n^\T\E[x_t],\] which is the mean sum of the state vector $x_t$. We first establish in Appendix \ref{append:spectra} that the operator norm of $\E[\A]$ is strictly less than 1 for $\beta \in(0,1)$. Hence, 
\[\lim_{t\to\infty}\E[\A]^t=0_{n\times n}\]
and
\[
\lim_{t\to\infty}\E[\A]^{t-1}+\cdots +I_n = (I_n-\E[\A])^{-1},
\]
which lead to the result that
\begin{align}\label{eq:y_infty_mat}
y_\infty:=\lim_{t\to\infty} \1_n^\T \E[x_t] = \1_n^\T (I_n-\E[\A])^{-1}\E[\B]x_0.
\end{align}
The first main result of this work is about the asymptotic mean sum of $x_t$. For given network Laplacian $L$, swapping rate $\alpha$, and the subset size $r$, we have for sufficiently small resetting rate $\beta$ that
\begin{align}\label{eq:y_infty_approx}
   y_\infty = \frac{n}{r}\sum_{i=1}^r x_0(i) + O(\beta).
\end{align}
The big $O$ notation is understood in the standard way. In general, the $O(\beta)$ term remains present and the small $\beta$ condition is essential for the 
approximation to hold. An exception is when $x_0(k)$ is constant for all $k\in[r]$, i.e., all designated nodes share the same initial value, in which 
case the $O(\beta)$ term vanishes exactly for any $\beta\in(0,1)$.
We show the detailed steps to derive Eq. (\ref{eq:y_infty_approx}) and discuss the special case in Appendix \ref{append:proof_y_infty}. It is worth noting that in such an expression for $y_\infty$, we highlight the role played by the resetting dynamics while keeping other factors fixed (such as the network structure as encoded in $L$ and the swapping rate $\alpha$). The exact expression for $y_\infty$ is given by Eq. (\ref{eq:y_infty}), and the dependence of $y_\infty$ on $\beta$ is explicitly shown in Eq. (\ref{eq:y_infty_beta}) to leading order in $\beta$. A sufficient condition under which $\beta$ can be considered small is
\begin{align}\label{eq:small_beta}
    \beta \left(\frac{1}{r}+\frac{m}{\alpha r}\frac{1}{\lambda_2}\right)\ll 1,
\end{align} 
where $\lambda_2$ is the spectral gap, or the second smallest eigenvalue of the network Laplacian $L$. The details to derive this condition are also given in Appendix~\ref{append:proof_y_infty}. In general, a ``small'' $\beta$ is easier to achieve in well connected networks (large $\lambda_2$) than in poorly connected ones. However, Eq.~(\ref{eq:y_infty_approx}) indicates that the influence of the network structure is of secondary importance. Specifically, when $\beta$ is fairly small, our result holds irrespective of the network details. This is verified by our numerical simulations below in Sec.~\ref{sec:experiments}. 

Equation (\ref{eq:y_infty_approx}) shows that the global state will eventually be influenced by a designated subset of nodes with resetting dynamics. For small $\beta$, the main contribution is just from the subset of selected nodes, and the contribution from all other nodes in the network is secondary and vanishes as $\beta\to 0$ by Eq. (\ref{eq:y_infty_approx}):
\[
\lim_{\beta\to 0}\lim_{t\to\infty} \1_n^\T \E[x_t] = \frac{n}{r}\sum_{i=1}^r x_0(i).
\]
Intuitively, the whole network seems to ``consist of'' $n/r$ copies of the subset of nodes when it comes to the state sum. However, if we let $\beta=0$ first, then the dynamics only contains the swapping process, which preserves the sum of $x_t$ at any $t$, i.e., for $\beta=0$, we have at any time $t$ that
\[
\1_n^\T x(t) = \1_n^\T x_0(t).
\]
The two different scenarios imply that the $t\to\infty$ and $\beta\to 0$ limits do not commute:
\[
\lim_{\beta\to 0} \lim_{t\to\infty} \1_n^\T \E[x_0] \neq \lim_{t\to\infty} \lim_{\beta\to 0} \1_n^\T \E[x_0].
\]
This noncommutativity of the limits arises because the resetting dynamics breaks the conservation of the value sum that holds under the swapping dynamics. If $\beta$ is first set to 0, then $\E[\B]$ becomes $0_{n\times n}$ and $\E[\A]$ reduces to $T_2=I_n -\alpha L/m$. The operator norm of $\E[\A]$ thus becomes 1, and the influence of the initial state will not decay over time. In fact, in this case, we have at any time $t$ that
\begin{align*}
    \E[\A]^t x_0 &= (I_n-\alpha L/m)^t x_0= \sum_{i=1}^n (1-\alpha\lambda_i/m)^tv_iv_i^\T x_0\\
    &= v_1v_1^\T x_0 +\sum_{i=2}^n(1-\alpha\lambda_i/m)^tv_iv_i^\T x_0\\
    &= \frac{\sum_{i=1}^n x_0(i)}{n}\1_n + \sum_{i=2}^n(1-\alpha\lambda_i/m)^tv_iv_i^\T x_0,
\end{align*}
where $0=\lambda_1<\lambda_2\leq\cdots\leq \lambda_n$ are eigenvalues of $L$ and $v_i$'s are the corresponding orthonormal eigenvectors. In particular, $v_1=\1_n/\sqrt{n}$ represents the steady-state mode of the dynamics. For all other modes, the corresponding coefficient $(1-\alpha\lambda_i/m)^t$ will converge to 0 as $t\to\infty$. (Detailed discussions on the spectra of $\E[\A]$ are given in Appendix \ref{append:spectra}.) Hence, the dynamics indeed preserves the sum of the state vector for $\beta=0$:
\[
\lim_{t\to\infty} \1_n^\T \E[\A]^t x_0 = \sum_{i=1}^n x_0(i).
\]
This is consistent with the understanding that swapping only (partially) exchanges the state at each node, and the sum should not be changed. Nonetheless, the introduction of resetting dynamics leads to a nontrivial effect, even if the rate $\beta$ is small. In the context of social influence, our result thus predicts that a group of persistent influencers, even if not directly connected, will reach consensus, with the opinions of others playing only a secondary role. While from the distributed computing perspective, this result paves the way for establishing the resetting mechanism as a tool for computing the mean of a subsample, which is our ultimate goal.

\subsection{Finite-time mean state and convergence rate}
\label{subsec:finit_time_convergence_rate}
A natural and highly relevant question for the application of our approach is how quickly the mean sum $y_t$ converges to  $y_\infty$. Our result is practically useful only if the convergence from the initial state sum  $y_0=\sum_{i=1}^n x_0(i)$ to the small-$\beta$ long-time limit $\frac{n}{r}\sum_{i=1}^r x_0(i)$ is sufficiently fast. We are thus interested in quantifying the difference between $y_t$ and $y_\infty$ at a finite $t$ (and for a small $\beta$). We denote the difference as
\[
\Delta_t := y_t - y_\infty.
\]
Clearly, $\Delta_t$ consists of two parts. The first part $\Delta_{t1}$ originates from the difference between $\1_n^\T\E[\A]^tx_0$ and its long-time limit:
\[
\Delta_{t1} = \1_n^\T \E[\A]^t x_0 - \underbrace{\lim_{t\to\infty}\1_n^\T \E[\A]^t x_0}_{=0} = \1_n^\T \E[\A]^t x_0.
\]
As discussed above, the long-time limit vanishes since the operator norm $\lVert \E[\A]\rVert<1$. Similarly, the other part $\Delta_{t2}$ is the difference between $\1_n^\T(\E[\A]^{t-1}+\cdots+I_n)\E[\B]x_0$ and its corresponding long-time limit $\1_n^\T(I_n-\E[\A])^{-1}\E[\B]x_0$. This difference is again closely related to the spectral property of $\E[\A]$. As the operator norm $\lVert \E[\A] \rVert $ is strictly less than 1, the convergence to the limiting state is exponentially fast. 

However, for operational purposes, an approximate but informative convergence rate might be more preferred. We obtain such a rate by first noticing that $\E[\A]$ almost commutes with $L$ at small $\beta$ and then using matrix perturbation theory. Details are given in Appendix \ref{append:convergence_rate}. Here, we report our second main result of this work: For large $t$ and small $\beta$, we have
\begin{align}\label{eq:convergence_rate}
    \Delta_t \approx \left(1-\frac{\beta}{n}\right)^t\Delta_0,
\end{align}
where the initial gap $\Delta_0$ is approximately \[\Delta_0\approx\sum_{i=1}^n x_0(i)-\frac{n}{r}\sum_{i=1}^r x_0(i).\] 
Roughly speaking, this result shows that the mean sum is changed from the initial value to the desired limit exponentially fast for any fixed small $\beta$. Therefore, for a given precision, $y_t$ converges sufficiently close to $y_\infty$ in finite time. 

Operationally, this result defines a critical time beyond which the absolute difference $|\Delta_t|$ is smaller than the given precision. That is, for some fixed $\varepsilon>0$, we have $|\Delta_t|\lesssim \varepsilon$, as long as 
\begin{align}\label{eq:exponential_decay}
    t\geq \frac{\ln(\varepsilon/|\Delta_0|)}{\ln(1-\beta/n)}.
\end{align}
Suppose we know $|\Delta_0|,\beta$, and $n$, then an estimate of how long the resetting dynamics should be applied can be calculated such that the finite-time mean state sum $y_t$ is close to its long-time limit $y_\infty$. Even if $|\Delta_0|$ is unknown, we may still estimate the critical time needed to achieve a prescribed relative error (for example, $\varepsilon=0.1|\Delta_0|$), provided that an upper bound of the network size $n$ and a lower bound of the resetting rate $\beta$ are available. 

\subsection{Probabilistic bounds for a single trajectory}
\label{subsec:probalisitc_bound_for_single_trail}
So far, we have analyzed the mean sum both in the long-time limit and at finite times. We have shown that, asymptotically, the mean sum is approximately driven to $\frac{n}{r}\sum_{i=1}^r x_0(i)$, with an average convergence rate of roughly $(1-\frac{\beta}{n})^t$. From a practical standpoint, however, there remains one important question: Since the process is random, what can we say about the behavior of a single realization of the process?

To this end, we now examine the behavior of a single random trajectory. We let \[Y_t:=\1_n^\T x_t\] be a random realization of the state sum at time $t$ and $y_t=\E[Y_t]$ is the expectation. We want to answer this question: How far is $Y_t$ from $y_t$? Operationally, if one can observe a single trajectory of the process, and can quantify the difference between $Y_t$ and $y_t$ (with high probability), then it is possible to control the process to a given precision $\varepsilon>0$. Our third main result is
\begin{align}\label{eq:hpb_Yt}
    \p\left(|Y_t - y_t|>\varepsilon\right) \lesssim 2\exp\left(-\frac{\varepsilon^2}{n\beta\R^2}\right),
\end{align}
where $\sigma_r = [\max_{k\in[r]}x_0(k) - \min_{k\in[r]}x_0(k)]/2$ is half the range of the node values in the designated subset. Technically, we establish that $Y_t$ is a sub-Gaussian random variable and nontrivially bound its tail probability using Hoeffding's inequality \cite{van2014probability}. This high-probability bound is based on an informative sub-Gaussian parameter associated with $Y_t$, which results from an effective dynamics of the process where the randomness of $Y_t$ is mainly contributed by $\B_\tau$ at each time step. We defer the detailed derivation of  (\ref{eq:hpb_Yt}) to Appendix \ref{append:confidence_interval}.

Combining (\ref{eq:exponential_decay}) and (\ref{eq:hpb_Yt}) together, an important corollary can be obtained that, for $\varepsilon>0$ and $\delta\in(0,1)$, we have
\begin{align}\label{eq:stat_guarantee}
    \p\left(|Y_t - y_\infty|\leq\varepsilon\right) \gtrsim 1-\delta
\end{align}
if $\beta\leq \beta^\ast$ and $t\geq t_\ast$, where 
\begin{align}\label{eq:parameters}
\beta^\ast = \frac{\varepsilon^2}{4n\ln(2/\delta)\sigma_r^2} \text{ and }t_\ast = \frac{\ln(\varepsilon/2|\Delta_0|)}{\ln(1-\beta/n)}.
\end{align}
This result is a consequence of the fact that
\begin{align*}
    &\p(|Y_t-y_\infty|>\varepsilon)\leq\p\left(|Y_t-y_t|>\frac{\varepsilon}{2}\right) + \p\left(|y_t-y_\infty|>\frac{\varepsilon}{2}\right).
\end{align*}
The first term on the right-hand side characterizes the intensity of the random fluctuation of $Y_t$ around its mean $y_t$, and can be bounded by (\ref{eq:hpb_Yt}). The second term is either 0 or 1, depending on whether the time is long enough so that the difference between $y_t$ and $y_\infty$ is sufficiently small. This term is controlled by (\ref{eq:exponential_decay}) and equals 0 for $t>t_\ast$. 

This result is of great relevance in protocol design. Practically, if the necessary information of $n,r,x_0$ is known, then we can set the values of $\beta$ and $t$ based on prescribed $\varepsilon$ and $\delta$, and the resulting state of a single stochastic trajectory will have a \emph{random} sum $Y_t$ close to the deterministic  limit $y_\infty$ with a statistical guarantee.

\section{Numerical experiments}
\label{sec:experiments}

In this section, we present systematic numerical experiments to validate our theoretical findings. By Eq.~(\ref{eq:parameters}): for a given precision $\varepsilon$ and miscoverage rate $\delta$, the resetting rate $\beta^\ast$ required to guarantee convergence depends on the network size $n$ 
and the half-range of the designated node values $\sigma_r$. The convergence time $t^*$ further depends on $\beta^\ast$ and the initial gap \(\Delta_0\).

Importantly, the convergence behavior under the parameters prescribed by our formula typically does not depend on the network topology or spectral properties, provided the small-$\beta$ condition (\ref{eq:small_beta}) holds.

Our theoretical results hold for general connected networks, and we design the numerical experiments to test the protocol across representative network structures. We use Erd\H{o}s-R\'enyi (ER), lattice, and Barab\'asi-Albert (BA) networks as canonical examples of homogeneous random, regular spatial, and heterogeneous hub-dominated topologies, respectively. In the controlled
ER experiments, we use connection probabilities \(p\in[0.3,0.5]\) to
obtain connected homogeneous random graphs, and vary one factor at a
time, such as the range of designated values, the size of the
designated set, the network size, or the initial sum gap \(\Delta_0\),
while all other factors are held fixed. In
Fig.~\ref{fig:different_networks}, we further include {a sparse ER network ($p=0.08$)} together with a lattice and a BA network, to compare the influence of different structural regimes and edge densities.

Throughout, $Y_t = \mathbf{1}_n^{\mathsf{T}} x_t$ denotes the state sum of a single random realization, while $y_t = \mathbb{E}[Y_t]$ denote its mean and $y_\infty = \lim_{t\to\infty} y_t$ denote the long-time limit of $y_t$.  This, alongside other key notation, is summarized in  Table~\ref{tab:notation}. Further, each of the following figures shows 50 independent realizations of the evolution of $Y_t$.

We begin with an illustrative example  that introduces the key metrics and demonstrates the core convergence result in Sec.~\ref{subsec:exp_example}. We then show the dynamical consequence when the resetting rate $\beta$ is large and falls outside the regime predicted by our formula in Sec.~\ref{subsec:exp_large_beta}. We next investigate the impact of different designated node configurations in Sec.~\ref{subsec:exp_different_configurations_designated_nodes}. In
Sec.~\ref{subsec:exp_network_size}, we study the effect of network size $n$ on convergence, and in Sec.~\ref{subsec:exp_different_nodes_value_distribution} we explore how the initial sum gap $\Delta_0$ shapes convergence behavior. Finally, we demonstrate in Sec.~\ref{subsec:exp_network_structures} that, under the prescribed parameters, the convergence behavior is universal across network topologies with distinct spectral gaps, confirming the result of Eq. (\ref{eq:y_infty_approx}) that the spectral gap's influence on $y_\infty$ via $\beta$ serves as a first order correction.

\begin{table}[t]
\caption{Notation used in the plots.}
\label{tab:notation}
\begin{ruledtabular}
\begin{tabular}{ll}
Symbol & Meaning \\
\colrule
$x_t(i)$ & State of node $i$ at time $t$ (random) \\
$Y_t = \mathbf{1}_n^{\mathsf{T}} x_t$ & State sum of a single random realization \\
$y_t = \mathbb{E}[Y_t]$ & Mean state sum \\
$y_\infty = \lim_{t\to\infty} y_t$ & Long-time limit of the mean sum \\
$\Delta_t = y_t - y_\infty$ & Gap to the limiting mean \\
\end{tabular}
\end{ruledtabular}
\end{table}

\subsection{Illustrative example}
\label{subsec:exp_example}
We first use a concrete example to show the main findings of this work. We construct an ER network \cite{erdds1959random} with $n=30$ and the connection probability for each pair of nodes is $p=0.5$. The index set for the designated nodes is $[r]$ with $r=5$. The initial values of $x_0(k),k\in[r]$ are drawn independently from the Gaussian distribution $N(-1,1^2)$ with mean $-1$ and variance $1^2$. Initial values for other nodes are independently drawn from $N(0,10^2)$. The swapping rate is $\alpha=0.5$. We numerically calculate $y_\infty$ and $y_t$ based on Eqs. (\ref{eq:y_infty_mat}) and (\ref{eq:Ext}), respectively. The process stops at $t=t_\ast$ and the resetting rate $\beta$ is set to be $\beta^\ast$. Both $t_\ast$ and $\beta^\ast$ are calculated based on Eq. (\ref{eq:parameters}) with precision $\varepsilon=1$ and miscoverage rate $\delta=0.1$. 

\begin{figure}
    \centering    
    \includegraphics[width=0.5\textwidth]{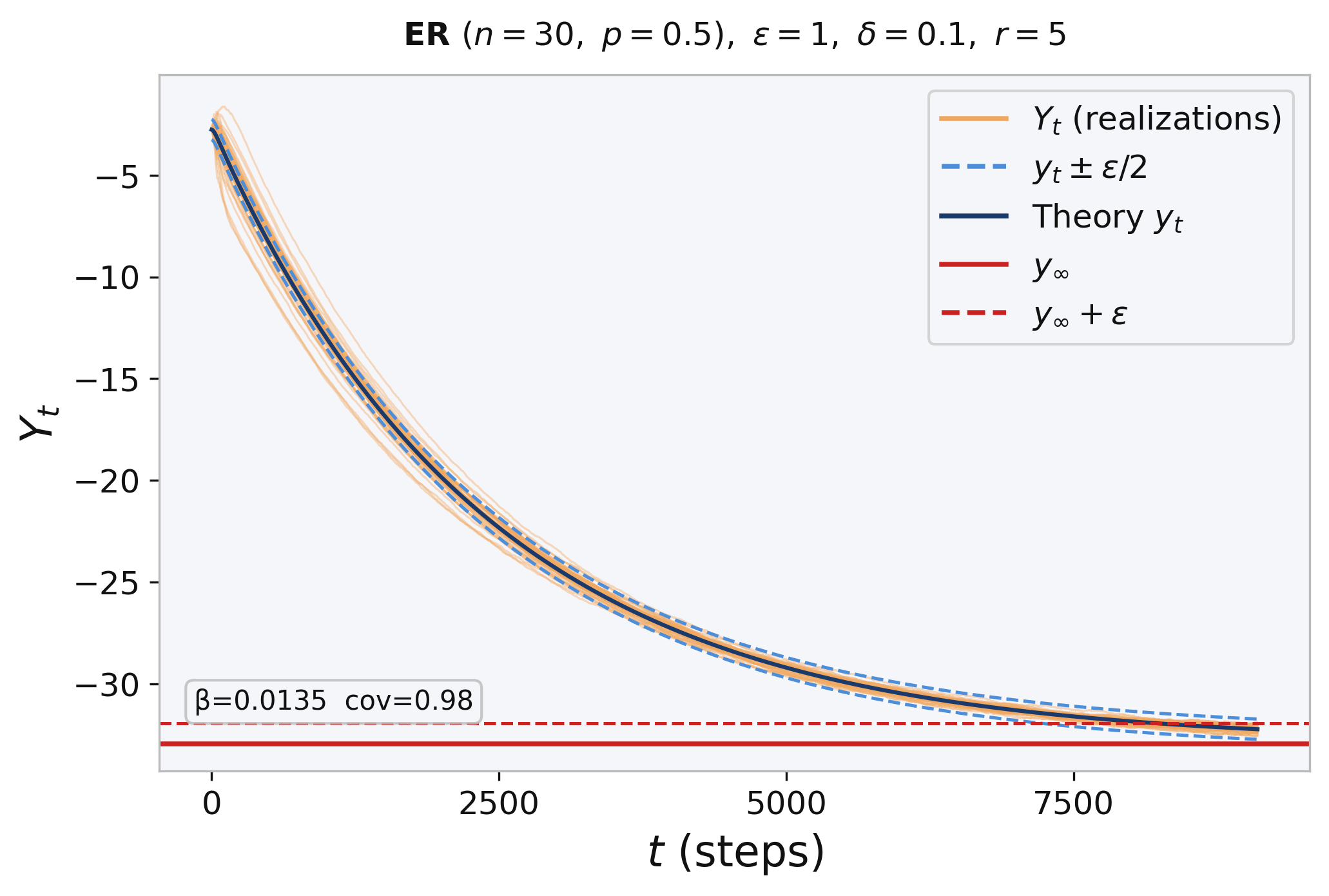}
    \caption{An ER network with $n=30$ and $p=0.5$. Nodes 1 through 5 undergo resetting dynamics with initial states independently and identically distributed (iid) as $x_0(1),\ldots,x_0(5)\overset{iid}{\sim}N(-1,1^2)$. The initial states for nodes with $i> 5$ are $x(i)\overset{iid}{\sim}N(0, 10^2)$. The two blue dashed lines correspond to $y_t\pm \varepsilon/2$, respectively (with $\varepsilon=1$). The orange lines represent 50 realizations of the evolution of $Y_t$. We use Eq. (\ref{eq:parameters}) to calculate the critical parameters $\beta^\ast$ and $t_\ast$, and then let $\beta=\beta^\ast$ be the resetting rate and stop the process at time $t_\ast$. Our theory guarantees that the coverage rate is approximately at least $1-\delta=0.9$. That is, with high probability (approximately greater than 90\%), a random realization of $Y_{t_\ast}$ will lie within the interval $y_\infty\pm \varepsilon$. For this specific experiment, we numerically observe that the empirical coverage rate is $\mathrm{cov}=98\%$.}
    \label{fig:example}
\end{figure}

In Fig. \ref{fig:example}, we plot $y_t$ as well as the band $y_t\pm\varepsilon/2$, $y_\infty$, and 50 independent realizations of the process from $t=0$ to $t_\ast$. Our theory predicts that when the process stops at $t=t_\ast$, at least roughly $90\%$ of the trajectories will end up within the interval $y_\infty \pm \varepsilon$. Note that the probability bound is derived from Hoeffding's inequality, which can be quite conservative. We define the empirical coverage rate as 
\[
\mathrm{cov} := \hat{P}(|Y_{t_\ast} - y_\infty| < \varepsilon),
\] 
where $\hat P$ denotes the empirical distribution. That is, $\mathrm{cov}$ is the proportion of independent runs whose terminal value $Y_{t_\ast}$ falls within $\varepsilon$ of $y_\infty$ at the stopping time $t_\ast$. Numerically, 98\% trajectories are within this interval at $t=t_\ast$ for this example.

\subsection{Influence of a large resetting rate $\beta\gg\beta^\ast$}\label{subsec:exp_large_beta}

\begin{figure*}
    \centering
    \includegraphics[width=1\linewidth]{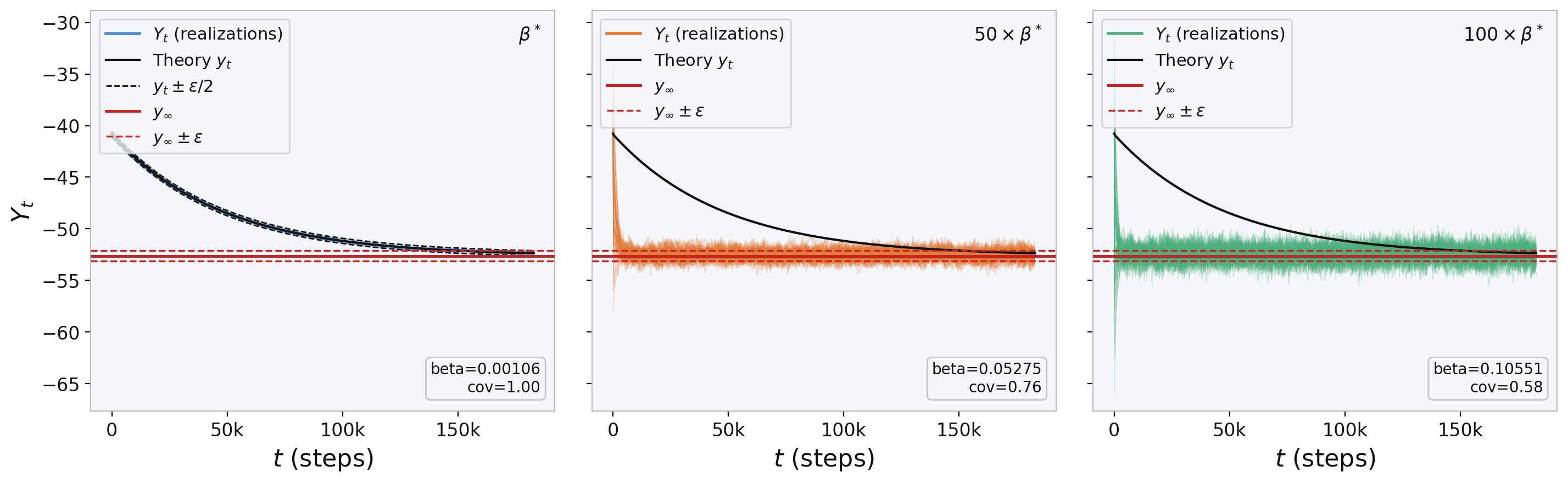}
    \caption{Effect of the resetting rate $\beta$ on the convergence behavior for an ER network under the same setting as in Fig.~\ref{fig:example}, except that $n = 50$. All three panels share the same network, initial conditions, and critical time horizon $t_\ast$ and resetting rate $\beta^\ast$. The solid black line shows the theoretical prediction $y_t$ under $\beta^\ast$, and colored lines show 50 individual simulation replicates. Left: $\beta = \beta^\ast$; the realizations $Y_t$ are tightly clustered around $y_t$, all replicates converge within $y_\infty \pm \varepsilon$ at $t_\ast$, yielding $\text{cov} = 1.00$. Center: $\beta = 50 \times \beta^\ast$; convergence is faster initially, but trajectories exhibit persistent fluctuations and coverage drops to $\text{cov} = 0.76$. Right: $\beta = 100 \times \beta^\ast$; fluctuations are even larger and coverage further drops to $\text{cov} = 0.58$. In the latter two cases, trajectories fail to remain within $y_\infty \pm \varepsilon$ even as $t \to \infty$.
    }
    \label{fig:varying_beta}
\end{figure*}

Intuitively, one may want to speed up the dynamics by adopting a larger $\beta$ than the theoretically predicted $\beta^\ast$ from Eq.~\eqref{eq:parameters}. However, if a substantially larger $\beta$ is used, the statistical guarantee \eqref{eq:stat_guarantee} will be lost. In Fig.~\ref{fig:varying_beta}, we plot how a large resetting rate $\beta$, which could be much larger than $\beta^\ast$, affects the dynamics. The settings are similar to those in Fig.~\ref{fig:example} except that $n=50$. All three panels use the same ER network, hence, the same $t_\ast$ and $\beta^\ast$. The first panel adopts $\beta^\ast$ as the resetting rate, the second uses $50\times\beta^\ast$, and the third uses $100\times\beta^\ast$. A larger $\beta$ accelerates the initial convergence speed: trajectories under $50\times\beta^\ast$ and $100\times\beta^\ast$ resetting rates reach the vicinity of $y_\infty$ much faster than under $\beta^\ast$. However, this comes at a cost. First, we observe that individual trajectories exhibit larger variations when $\beta$ is large, as frequent resetting introduces more noise into the dynamics. Second, and more critically, the trajectories fail to settle. Instead, they maintain persistent fluctuations around $y_\infty$, with the amplitude of oscillations growing with $\beta$. As a result, coverage degrades significantly, dropping to $0.76$ at $50\times\beta^\ast$ and further to $0.58$ at $100\times\beta^\ast$, and the trajectories fail to remain within $y_\infty \pm \varepsilon$ even as 
$t \to \infty$. 

In fact, once the network structure, the designated set, initial values, precision $\varepsilon$, and miscoverage rate $\delta$ are given, $\beta^\ast$ is determined, and $t_\ast$ is also consequently determined. Increasing $\beta$ will drive the finite-time mean sum $y_t$ to $y_\infty$ faster, but leaving the fluctuation of a random $Y_t$ around $y_t$ uncontrolled. Hence, there is a tradeoff between how fast the mean dynamics converges to its limit and how tightly a finite-time realization concentrates around its mean. The results in Fig.~\ref{fig:varying_beta} confirm that, when $\varepsilon$ and $\delta$ are prescribed, $\beta^\ast$ achieves a balance between bias (the deterministic difference $|y_t-y_\infty|$) and variance (characterized by the random fluctuation $|Y_t-y_t|$) and it plays an essential role in ensuring both convergence and stability of our protocol.

\subsection{Influence of designated nodes}
\label{subsec:exp_different_configurations_designated_nodes}

We explore the effects of different configurations of the designated nodes below, including the range of their initial values, the number of designated nodes, and their degrees.

\subsubsection{Range of initial values}

\begin{figure}
    \centering
    \includegraphics[width=1\linewidth]{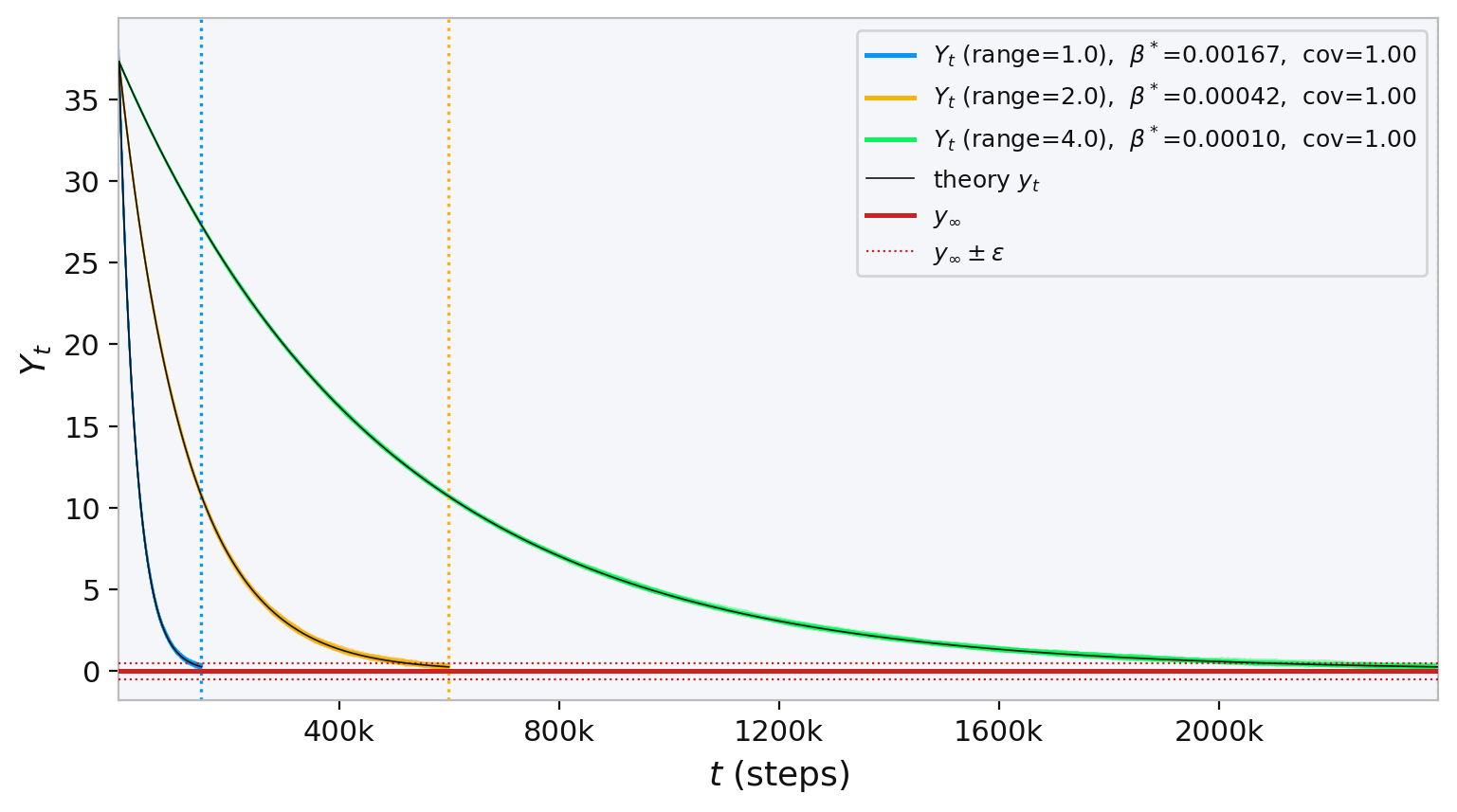}
    \caption{Effect of the range of designated node initial values on the  convergence behavior. The setup is the same as in Fig.~\ref{fig:varying_beta}, except for the initial values of the nodes in the designated set. We consider three sets of such initial values with the same mean $0$ but different ranges: $\{-0.5, -0.3, 0, 0.3, 0.5\}$ (tight), $\{-1.0, -0.5, 0, 0.5, 1.0\}$ (medium), and $\{-2.0, -0.7, 0, 0.7, 2.0\}$ (wide). Note that $y_\infty$ is also the same for all cases. As the range widens, $\beta^\ast$ decreases and $t_\ast$ increases (vertical dotted lines) according to our theory. Numerically, the coverage rate $\mathrm{cov} = 1.00$ is achieved at $t_\ast$ in all three cases.}
    \label{fig:different_desi_range}
\end{figure}

We examine how the range of designated node initial values affects convergence behavior in Fig.~\ref{fig:different_desi_range}. The setup is the same as that in Fig.~\ref{fig:varying_beta}, except for the initial values of designated nodes. We consider three sets of such values,  all with the same mean, so the asymptote $y_\infty$ is identical across all three cases. However, we consider three different ranges: $1.0$ (tight), $2.0$ (medium), and $4.0$ (wide). Since $\beta^\ast$ depends inversely on the squared range of $\{x_0(k)\}_{k\in[r]}$ [Eq.~\eqref{eq:parameters}], a wider range yields a smaller $\beta^\ast$ and consequently a longer stopping time $t_\ast$. This is reflected in Fig.~\ref{fig:different_desi_range}: for the tight range case, the convergence rate is the highest ($\beta^\ast = 0.00167$), followed by the medium ($\beta^\ast = 0.00042$) and wide ($\beta^\ast = 0.00010$) range cases. Theoretical stopping times $t_\ast$ are marked by vertical dotted lines. Despite these differences, all three settings achieve $\mathrm{cov} = 1.00$ at $t_\ast$, confirming the statistical guarantee of our theory.

\subsubsection{Number of designated nodes}
\label{subsec:number_of_designated_nodes}
\begin{figure}
    \centering
    \includegraphics[width=1\linewidth]{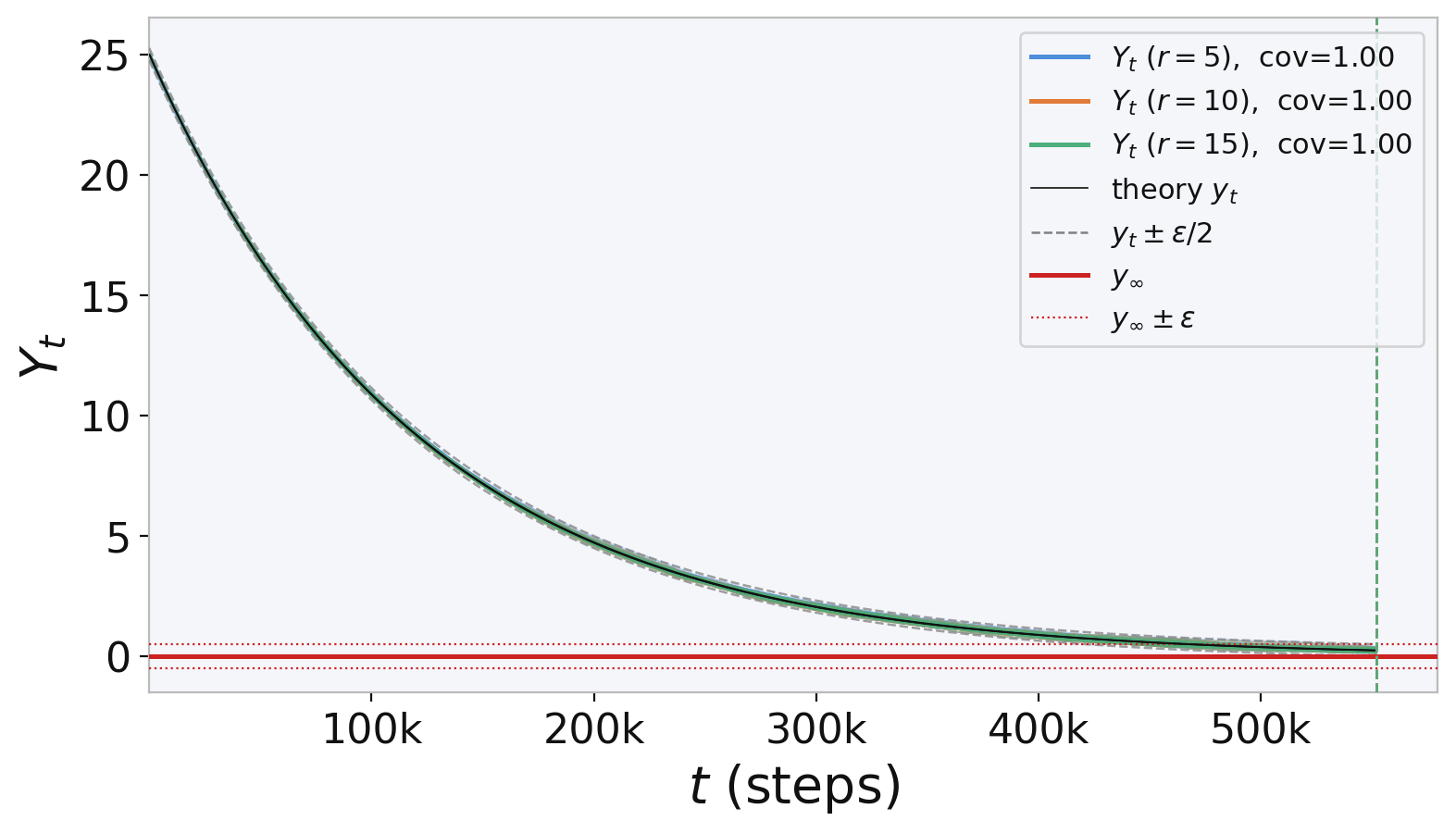}
    \caption{Effect of the number of designated nodes $r$ on the convergence behavior (ER network, $n=50$, $p=0.3$, $\alpha=0.5$, $\varepsilon=0.5$, $\delta=0.1$, 50 replicates). Initial values for designated node are evenly spaced in $[-1, 1]$ with mean zero for all $r\in\{5,10,15\}$; initial values for other nodes are rescaled so that the sum of initial values $\sum_{i=1}^n x_0(i)$, and hence $y_\infty$, is identical across all three cases. The theoretical curve $y_t$ and its $\pm\varepsilon/2$ band are plotted for each $r$; faint lines are individual replicates. However, the difference between them is too small, and they are visually indistinguishable, almost collapsing onto a single curve. In all three cases, we find $\mathrm{cov} = 1.00$ at their respective $t_\ast$, confirming that the coverage guarantee holds for different choices of $r$.}
    \label{fig:varying_designated_number}
\end{figure}
\begin{figure*}
    \centering
    \includegraphics[width=1\linewidth]{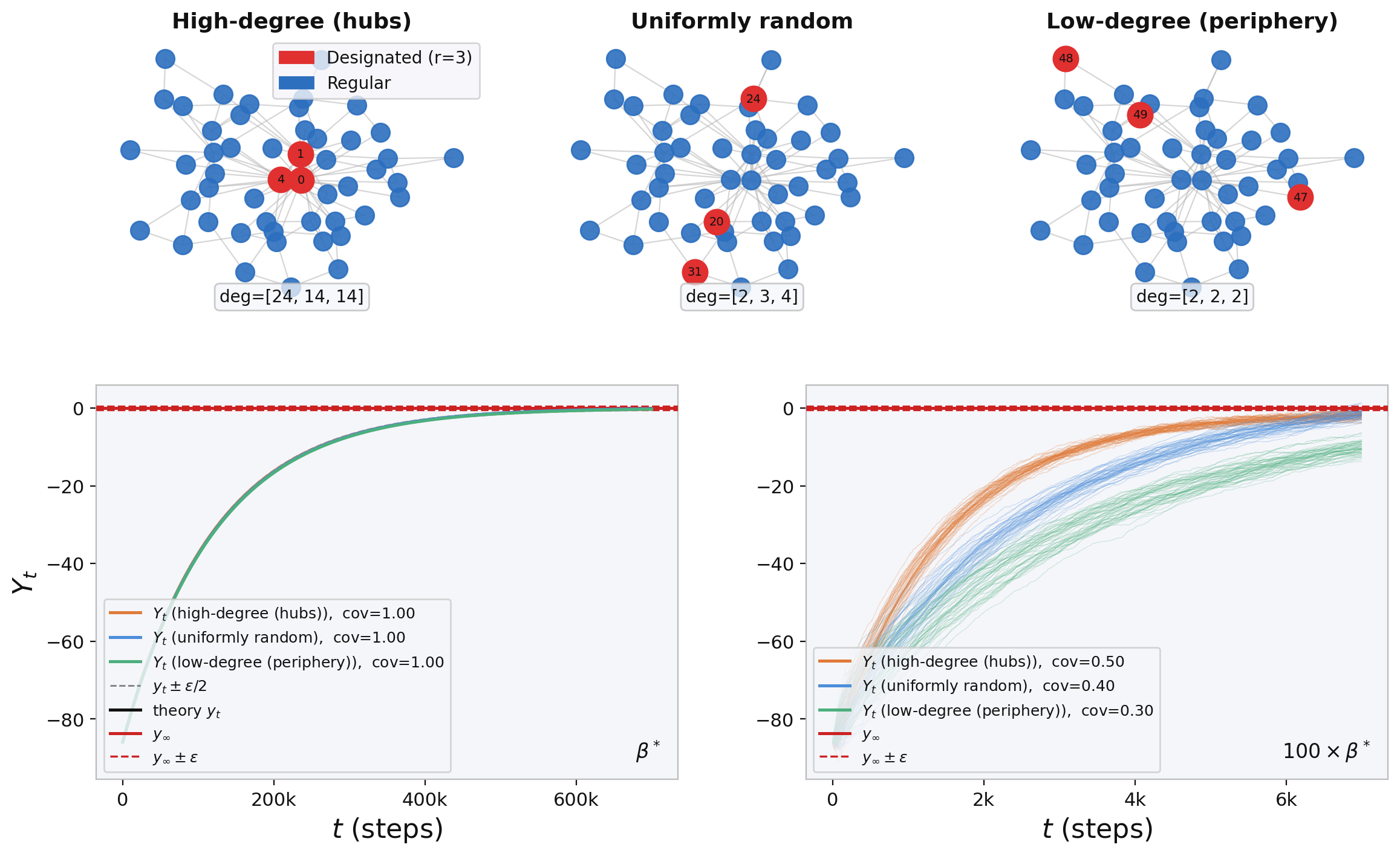}
    \caption{Effect of designated node selection on convergence behavior (BA scale-free network, $n=50$, $r=3$, $\alpha=0.5$, $\varepsilon=0.5$, $\delta=0.1$). Top row: Network visualizations for the three selection strategies: hubs (deg $= [24, 14, 14]$), random (deg $= [2, 3, 4]$), and periphery (deg $= [2, 2, 2]$). Bottom row: Mean trajectories (bold) and individual replicates (faint) over 50 runs, under $\beta = \beta^\ast$ (left) and $\beta = 100 \times \beta^\ast$ (right). Under $\beta = \beta^\ast$, all three strategies achieve $\mathrm{cov} = 1.00$ at $t_\ast$. Under $\beta = 100 \times \beta^\ast$, the hub strategy converges noticeably faster, while all strategies exhibit increased trajectory variance and oscillation, with no convergence guarantee.}
    \label{fig:different_desi_degree}
\end{figure*}
We in Fig.~\ref{fig:varying_designated_number} examine how the number of designated nodes affects the convergence behavior. We use an ER network for demonstration with three choices of $r \in \{5, 10, 15\}$. 

To evaluate the impact of $r$, we apply two control strategies in the experiment to keep $\sigma_r$ and $\Delta_0$ fixed. (1) To fix $\sigma_r$, the designated node initial values are always evenly spaced within $[-1, 1]$. For example, for $r=5$ the initial values are $\{-1, -0.5, 0, 0.5, 1\}$. In this case, the half-range $\sigma_r = 1$ for all $r$, which, by Eq.~(\ref{eq:parameters}), keeps $\beta^\ast$ identical across the three cases. Moreover, $\sum_{k=1}^r x_0(k) = 0$ and thus a common target $y_\infty \approx 0$ for all $r$. (2) To fix $\Delta_0 = \sum_{i=1}^n x_0(i) - \frac{n}{r}\sum_{i=1}^r x_0(i)$, it suffices to fix the total initial sum $\sum_{i=1}^n x_0(i)$. For $r=5$, the initial values for the $n-5$ nondesignated nodes are iid draws from $N(0, 10^2)$ and their sum is recorded as $S$. For $r\in\{10,15\}$, a fresh set of $n-r$ values $\{\tilde g(j)\}_{j\ge r+1}^n$ is independently drawn from the same distribution and then rescaled multiplicatively as the nondesignated node initial values: $x_0(j) = \tilde g(j) \times S / \sum_k \tilde g(k)$. One readily finds that they sum exactly to $S$. Since the designated values sum to zero in every case, the total sum $\sum_{i=1}^n x_0(i) = S$, and hence the initial gap $\Delta_0$ is identical for all $r$, which further leads to the same $t_\ast$ by Eq.~(\ref{eq:parameters}). With $\sigma_r$ and $\Delta_0$ now both fixed, all three cases share the same $\beta^\ast$ and $t_\ast$, so that any observed difference in the convergence behavior is due to $r$ alone.

As shown in Fig.~\ref{fig:varying_designated_number}, all three trajectories are nearly indistinguishable and all achieve $\mathrm{cov} = 1.00$ at their respective $t_\ast$. Thus, with $\beta = \beta^\ast$, the convergence guarantee holds across different numbers of designated nodes.

\subsubsection{Degrees of designated nodes}
\label{subsec:degrees_of_designated_nodes}
We investigate how the selection of designated nodes affects the convergence behavior in Fig.~\ref{fig:different_desi_degree}. To make the distinction more pronounced, we use a Barab\'asi-Albert (BA) scale-free network~\cite{barabasi1999emergence} with $n=50$ and $r=3$, and pick the designated set in three ways: (i) the $r$ nodes are those with the highest degrees (hubs), (ii) the $r$ nodes are chosen uniformly at random, and (iii) the $r$ nodes are those with the lowest degrees (periphery). The corresponding degree sequences for each case are shown in the top row of Fig.~\ref{fig:different_desi_degree}. 

In all three cases, the designated nodes are assigned the same initial values $\{-1, 0, 1\}$, and the $n-r$ remaining node values are drawn once from $N(0, 10^2)$ and reused across the three cases. Consequently, the three cases use the same set of initial values, so the target $y_\infty$, the half-range $\sigma_r$, and the initial gap $\Delta_0$ are identical. By Eq.~(\ref{eq:parameters}), they share the same $\beta^\ast$ and $t_\ast$, and differ only in the choice of the designated set, whose nodes may have quite different degrees.

As shown in the left panel of the bottom row of Fig.~\ref{fig:different_desi_degree}, under $\beta = \beta^\ast$, all three strategies yield $\mathrm{cov} = 1.00$ at $t_\ast$, and their trajectories are essentially indistinguishable. This result shows that the convergence guarantee by our protocol does not depend much on the specific way to select the designated nodes, and holds regardless of whether hubs, randomly chosen nodes, or peripheral nodes are selected.
Subset selection may, however, influence the dynamics when $\beta$ is large, as shown in the right panel of the bottom row of Fig.~\ref{fig:different_desi_degree}. Under $\beta = 100 \times \beta^\ast$, the hub strategy converges faster than the random and periphery strategies. Nevertheless, as also shown in Fig.~\ref{fig:varying_beta}, large resetting rates increase trajectory variation and induce oscillations, and in general there is no convergence guarantee for fixed $\varepsilon$ and $\delta$.

\subsection{Influence of network size}
\label{subsec:exp_network_size}
\begin{figure}
    \centering
    \includegraphics[width=1\linewidth]{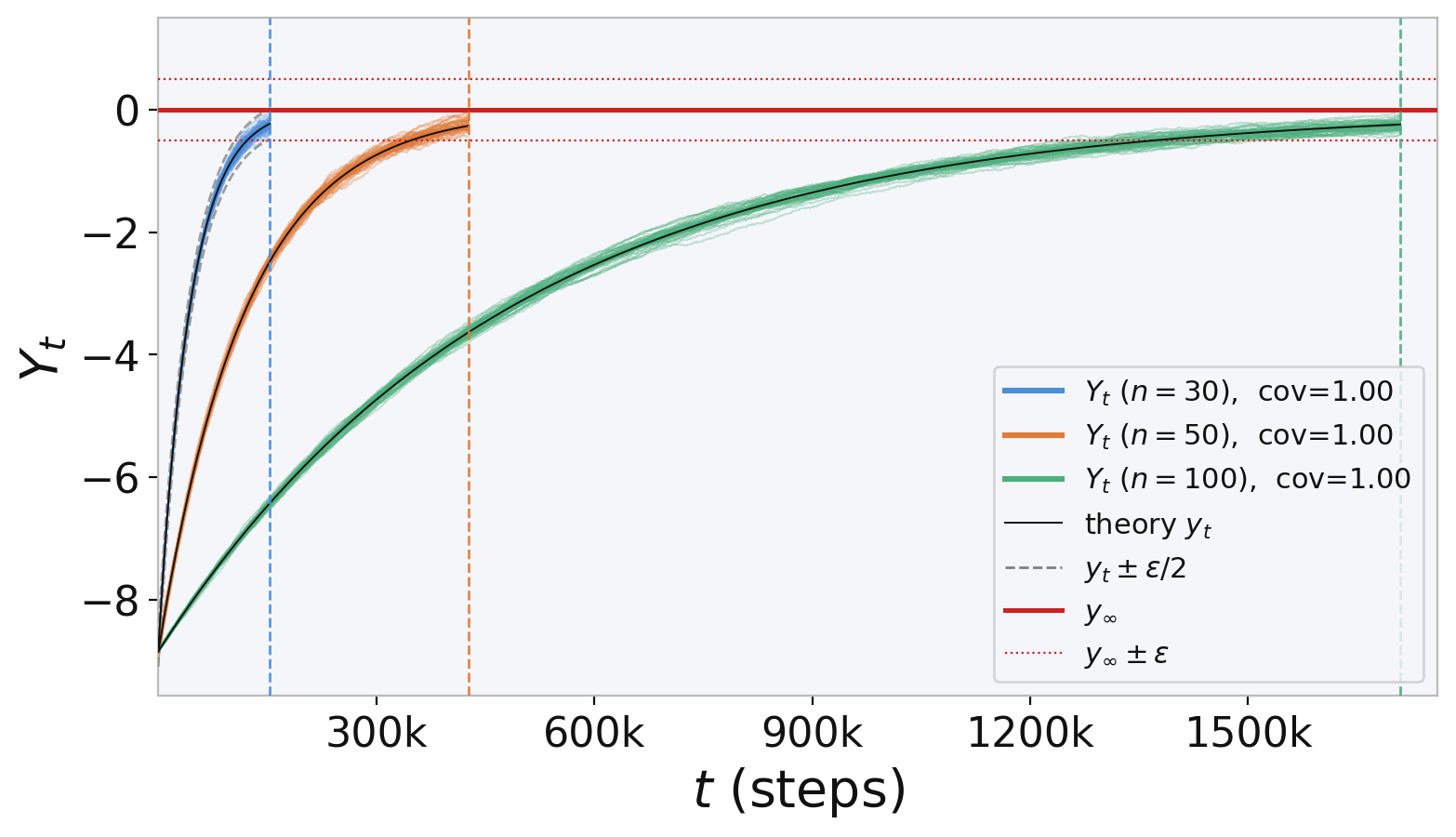}
    \caption{Effect of network size $n$ on convergence behavior (ER network, $p=0.3$, $r=5$, $\alpha=0.5$, $\varepsilon=0.5$, $\delta=0.1$, 50 replicates). Designated node values are fixed at $\{-1, -0.5, 0, 0.5, 1\}$; regular node values are rescaled so that their sum is identical across all three cases, keeping the initial gap $\Delta_0$ constant while $y_\infty \approx  0$. Larger networks require longer stopping times $t^*$ (vertical dashed lines), yet all three settings achieve $\mathrm{cov} = 1.00$ at their respective $t^*$.}
    \label{fig:different_network_size}
\end{figure}

We investigate how the network size affects the convergence behavior in Fig.~\ref{fig:different_network_size}. We fix the network type (ER, $p=0.3$) and the number of designated nodes ($r=5$), and vary the network size $n \in \{30, 50, 100\}$. To isolate the effect of $n$, we apply two controls. (1) The designated values $\{-1, -0.5, 0, 0.5, 1\}$ are the same for all $n$. Since they sum to zero, the target $y_\infty\approx \frac{n}{r}\sum_{k=1}^r x_0(k) = 0$ for every $n$, and their half-range $\sigma_r = 1$ is fixed. (2) For each $n$, the non-designated node initial values are drawn independently from $N(0, 10^2)$ and then rescaled so that $\sum_{i=1}^n x_0(i)$ matches the value used in the $n=30$ case. Consequently, the initial gap $\Delta_0$ is identical for all three cases. By Eq.~(\ref{eq:parameters}), $\beta^\ast$ and $t_\ast$ depend on the initial values only through $y_\infty$, $\sigma_r$, and $\Delta_0$. With these fixed, the network size $n$ is the only quantity that varies, and $\beta^\ast = \beta^\ast(n)$ and $t_\ast = t_\ast(n)$ depend on the corresponding $n$ in each case.
 
As shown in Fig.~\ref{fig:different_network_size}, larger networks converge more slowly: the stopping time $t^*$ increases substantially with $n$, as marked by the vertical dashed lines. This is expected since $\beta^\ast(n)$ decreases with $n$, and larger networks require more steps for the gossip mixing dynamics to propagate information across the graph. In fact, by Eq.~\eqref{eq:parameters}, $\beta^\ast$ is proportional to $n^{-1}$, and $t_\ast\sim n^{2}$. Due to the computational cost, we do not further increase  $n$ in our simulations. Nevertheless, all three settings considered here achieve $\mathrm{cov} = 1.00$ at their respective $t^*$, providing evidence that the convergence guarantee scales correctly with network size.

\subsection{Influence of initial sum gap $\Delta_0$}
\label{subsec:exp_different_nodes_value_distribution}
\begin{figure}
    \centering
    \includegraphics[width=1\linewidth]{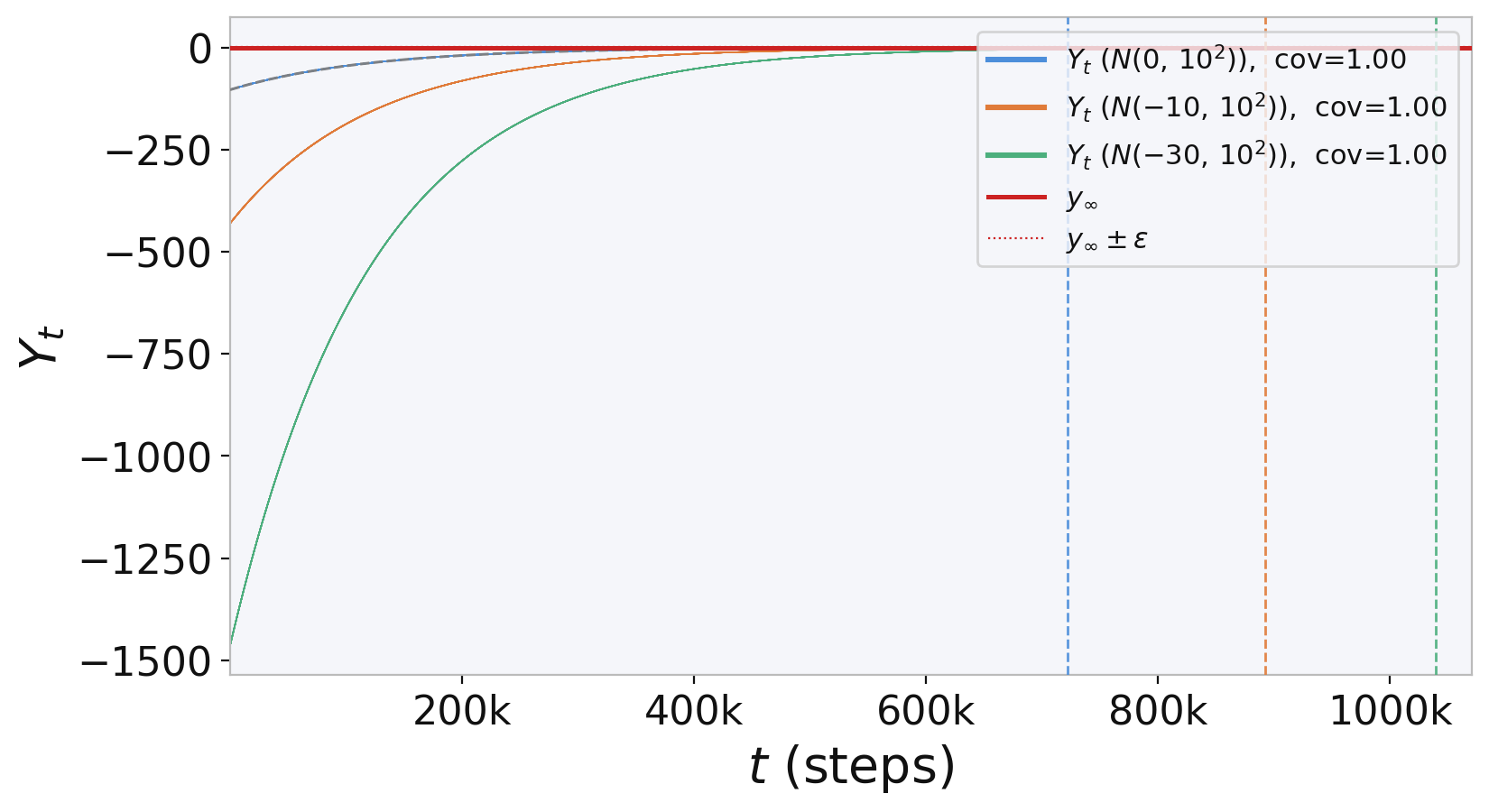}
    \caption{Effect of the initial sum gap $\Delta_0=y_0 - y_\infty$ on convergence behavior (ER network, $n=50$, $p=0.3$, $r=5$, $\alpha=0.5$, $\varepsilon=0.5$, $\delta=0.1$, 50 replicates). The network, designated nodes, and their initial values $\{-1, -0.5, 0, 0.5, 1\}$ are identical across all three cases. The initial values for other regular nodes are drawn from three Gaussian distributions with different means. A larger $|\Delta_0|$ results in a longer stopping time $t_\ast$ (vertical dashed lines). All three cases achieve $\mathrm{cov} = 1.00$ at their respective $t_\ast$.} \label{fig:different_network_node_value_range}
\end{figure}

Here we investigate how the initial sum gap $\Delta_0 = y_0 - y_\infty$ affects the convergence behavior. Specifically, we fix an ER network ($n=50$, $p=0.3$) with $r=5$ designated nodes with initial values $\{-1, -0.5, 0, 0.5, 1\}$ (note that the sum is 0), so that $y_\infty \approx 0$ in all cases. We then vary the mean of the Gaussian distribution from which the other $n-r$ regular node initial values are drawn: ${N}(0, 10^2)$, ${N}(-10, 10^2)$, and ${N}(-30, 10^2)$, respectively. As the Gaussian mean decreases, the corresponding gap $|\Delta_0|$ increases. By Eq.~\eqref{eq:parameters}, a longer stopping time $t_\ast$ is expected. We confirm this result in Fig.~\ref{fig:different_network_node_value_range}. The case with ${N}(0, 10^2)$ starts closest to $y_\infty$ and converges earliest, while the case with ${N}(-30, 10^2)$ starts furthest away from $y_\infty$ and requires the longest time to enter the $\varepsilon$ band. The stopping time $t_\ast$ in each case is marked by a vertical dashed line. Despite these differences in $|\Delta_0|$, all three cases achieve $\mathrm{cov} = 1.00$ at their respective $t_\ast$, confirming that the convergence guarantee holds across a wide range of initial sum gaps.

\subsection{Influence of network structure}
\label{subsec:exp_network_structures}
\begin{figure*}
    \centering
    \includegraphics[width=1\linewidth]{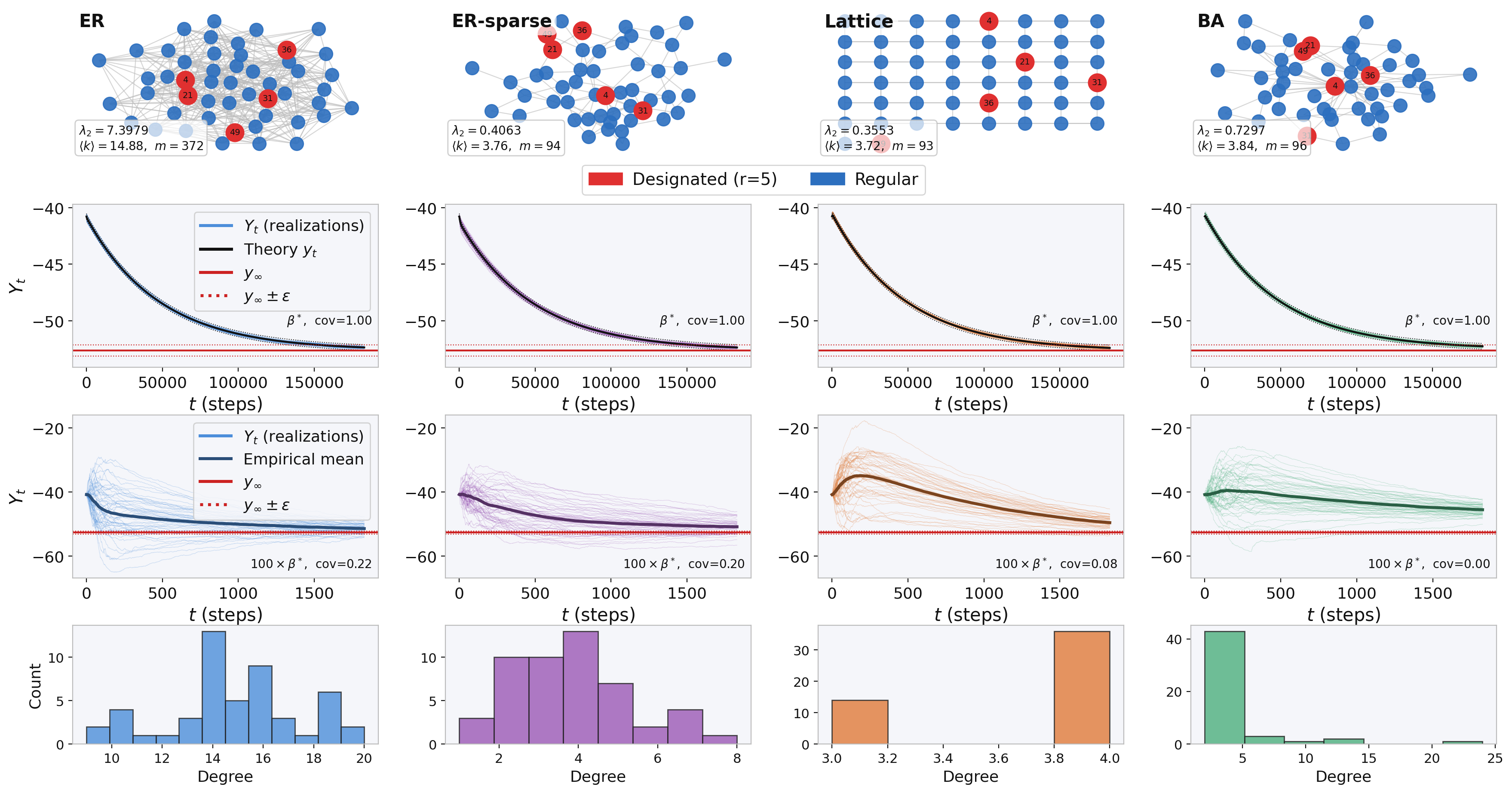}
    \caption{Comparison of gossip-with-reset dynamics across three network models: Erd\H{o}s-R\'enyi (ER), periodic lattice, and Barab\'asi-Albert (BA). The ER model is included at two densities, the original $p=0.3$ and a density-matched sparse variant (ER-sparse, $p=0.08$), giving four network instances in total, with $n=50$ nodes and $r=5$ designated nodes selected uniformly at random. The designated node initial states are drawn iid from $N(-1, 1^2)$ and those for the remaining nodes from $N(0, 10^2)$. The four networks have distinct spectral gaps $\lambda_2 = 7.3979$ (ER), $0.4063$ (ER-sparse), $0.3553$ (lattice), and $0.7297$ (BA). The sparse ER, lattice, and BA networks are density matched (mean degree $3.7$--$3.8$) while differing in degree heterogeneity and spectral gap, whereas the two ER networks share the same random-graph model but differ in edge density by roughly a factor of four; the comparison thus disentangles the effect of density from that of topology. Second row: Under $\beta = \beta^\ast$, all four networks achieve $\mathrm{cov} = 1.00$ at $t_\ast$, consistent with our convergence guarantee being insensitive to network topology under the small-$\beta$ condition~[Eq.~(\ref{eq:small_beta})]. Third row: Under large $\beta$ ($\beta = 100\times\beta^\ast$), the coverage guarantee no longer holds, and a large $\beta$ introduces significant variation among individual trajectories. Bold lines show the empirical mean, averaged across all Monte Carlo replicates at each time step. Fourth row: Degree distributions of the four network instances, illustrating the structural differences between ER (approximately Gaussian), lattice (near uniform), and BA (heavy tailed, hub dominated).}
    \label{fig:different_networks}
\end{figure*}
We compare the gossip-with-reset dynamics across three canonical network models: the ER random graph, the two-dimensional periodic lattice, and the BA scale-free network. To separate the effect of edge density from that of topology, we consider the ER model at two densities: one with $p=0.3$ as above (mean degree $\langle k\rangle \approx 14.9$) and a sparse variant (ER-sparse) with $p=0.08$ whose mean degree $\langle k\rangle \approx 3.8$ is matched to that of the lattice ($\langle k\rangle \approx 3.7$) and the BA ($\langle k\rangle \approx 3.8$) network. This yields four network instances in total.

All four networks have $n=50$ nodes, with $r=5$ designated nodes selected uniformly at random. The initial values of the designated nodes are iid draws from $N(-1, 1^2)$, and those of the remaining nodes from $N(0, 10^2)$. We run 50 independent replicates for each network. The four networks differ substantially in structure and spectral gap: $\lambda_2 = 7.3979$ for ER, $\lambda_2 = 0.4063$ for ER-sparse, $\lambda_2 = 0.3553$ for the lattice, and $\lambda_2 = 0.7297$ for BA. Specifically, ER is homogeneously well connected, the lattice has a rigid regular structure, and BA exhibits a highly heterogeneous degree distribution dominated by a few high-degree hubs (first row of Fig.~\ref{fig:different_networks}). The ER-sparse, lattice, and BA networks are density matched (mean degree $3.7$--$3.8$) but differ in degree heterogeneity and spectral gap, whereas the two ER networks share the same random-graph model and differ in edge density by roughly a factor of 4; the comparison therefore separates the effect of density from that of topology.

Despite these structural differences, when $\beta^\ast$ and $t_\ast$ are computed from our theoretical formula, all four networks achieve $100\%$ empirical coverage at the stopping time $t_\ast$ (second row of Fig.~\ref{fig:different_networks}). This provides evidence that, for the representative topologies considered here, our convergence guarantee is insensitive to the network topology, without requiring any prior knowledge of the network topology. In particular, the guarantee does not depend on the spectral gap $\lambda_2$, provided that the small-$\beta$ condition [Eq.~\eqref{eq:small_beta}] is satisfied. This is a notable departure from many classical gossip and diffusion dynamics, where the spectral gap directly governs the convergence rate and different network topologies would require different parameter choices.

The behavior changes, however, when $\beta$ exceeds $\beta^\ast$, as shown in the third row of Fig.~\ref{fig:different_networks} with $\beta = 100 \times \beta^\ast$. In this regime the coverage guarantee no longer holds for any of the four networks: the empirical coverage is $\mathrm{cov} = 0.22$ (ER), $0.20$ (ER-sparse), $0.08$ (lattice), and $0.00$ (BA), all far below the target $1-\delta = 0.9$, and individual trajectories exhibit substantial variation across all four topologies. This further illustrates that operating far beyond $\beta^\ast$ breaks the convergence guarantee regardless of network structure.

\section{Further discussion}\label{sec:discussion}

We mainly consider the task of subset averaging above, and one important ingredient of our protocol is calculating the ``weighted'' sum $\sum_{k=1}^r w(k)x_0(k)$ with $w(k)=1/r$. This can be generalized to the case that $w(k)$ is not constant. Concretely, if we let the resetting rate $\beta$ be \emph{node specific} such that $\beta(k)=\beta_0 w(k)$ for $k\in[r]$, with $\beta_0$ being a common factor, $w(k)\ge 0$, and $\sum_{k=1}^r w(k)=1$, then our analysis can still be applied with minor modifications. For example, the expression for $T_1$ in Eq.~(\ref{eq:EA}) is now $I_n-\beta_0 UWU^\T$ with $W=\text{diag}\{w(1),\ldots,w(r)\}$. Following similar steps in Appendix~\ref{append:proof_y_infty}, we arrive at $y_\infty = n \sum_{k=1}^r w(k)x_0(k) + O(\beta_0)$, which naturally reduces to (\ref{eq:y_infty_approx}) when $w(k)=1/r$. 

Our protocol may also be used to provide an approximate value of $\max_{k\in[r]}x_0(k)$, which is a highly nonlinear function of $\{x_0(k)\}_{k\in[r]}$. To this end, we note the relation that \[\max_k e^{\eta x_0(k)} \leq \sum_{k=1}^r e^{\eta x_0(k)}\leq r \max_k e^{\eta x_0(k)}\] for any $\eta>0$, which immediately implies 
\[
\max_k x_0(k)\leq \frac{1}{\eta}\ln
\left( \sum_{k=1}^r e^{\eta x_0(k)}\right) \leq \max_k x_0(k) + \frac{\ln r}{\eta}.
\] 
Hence, if a large $\eta$ is given, and our protocol is applied to a set of modified initial values $e^{\eta x_0(k)}$, then $\max_k x_0(k)$ can be approximated by a simple transformation of the protocol output.

\section{Conclusion}\label{sec:conclusion}
In this work, we theoretically analyze the effect of resetting (at rate $\beta$) on the gossip dynamics in a general connected network. We first observe that the long-time and small-$\beta$ limits do not commute. As long as the resetting mechanism is present, even if $\beta\ll 1$, the network will asymptotically converge to a state that is predominantly determined by a designated subset of nodes subject to resetting [Eq. (\ref{eq:y_infty_approx})]. These nodes may be scattered throughout the network and need not form a cluster geometrically. For small $\beta$,  we then show that the system converges exponentially fast to the asymptotic state and provides an approximate but informative rate [Eq. (\ref{eq:convergence_rate})]. Finally, we provide an approximate confidence interval for the mean sum of the system state at time $t$ [Eq. (\ref{eq:hpb_Yt})]. Combining these findings, we are able to identify the critical resetting rate and system evolution time [Eq. (\ref{eq:parameters})] such that, with high probability, the resulting random sum of the state vector is within a specified tolerance of the mean sum of the asymptotic state [Eq. (\ref{eq:stat_guarantee})].

It is worth noting that the resetting dynamics is \emph{local} as it involves only one node, the swapping dynamics is quasilocal as it involves two ``adjacent'' nodes, but the influence of the two types of dynamics is global. From a distributed systems perspective, the combination of resetting and swapping offers a means of controlling the network without relying on a central coordinator. This is beneficial not only for classical tasks, but also in quantum settings, where the resetting and swapping gates as natural generalizations of their classical counterparts can be applied to drive a quantum network toward some desired state determined by a subsystem. We hope the theory developed in this work could serve as a foundation for exploring such quantum applications, which may be crucial for advancing quantum communications.

\acknowledgements

X.M. and I.G. were supported in part by NSF CNS Grant No. 2504595, grants from the IBM-Illinois Discovery Accelerator Institute (IIDAI), and a gift from Microsoft. 

\begin{appendix}
\section{The operator norm of $\E[\A]$}\label{append:spectra}
Let $M$ be a generic $n\times n$ real matrix. In this section, we use $\lVert M\rVert$ to denote the operator norm of $M$. That is, $\lVert M\rVert := \max_{v\in\mathbb{R}^n:\lVert v\rVert=1}\lVert Mv\rVert$. Moreover, if $M$ has a spectral decomposition, then we use $\lambda_i(M)$ to denote the $i$th smallest eigenvalue of $M$ and let $v_i(M)$ be a corresponding orthonormal eigenvector.
\begin{lemma}\label{lem:spectra_EA}
We have $\lVert \E[\A] \rVert<1$.
\end{lemma}
\begin{proof}[Proof of Lemma \ref{lem:spectra_EA}]
    First note that $\E[\A]=T_1T_2$. It is obvious that $T_1$ as a diagonal matrix has eigenvalues $1-\beta/r$ and 1, and thus $\lVert T_1 \rVert = 1$. Also, $T_2$ commutes with $L$ and has eigenvalues
    \[
    \lambda_i(T_2) = 1-\frac{\alpha}{m}\lambda_{n+1-i}(L)
    \]
    with $0=\lambda_1(L)<\lambda_2(L)\leq \cdots\leq\lambda_n(L)$. Recall that since we consider a connected network, there is only one eigenvalue of $L$ that is exactly equal to 0. The maximal eigenvalue $\lambda_n(L)$ can be upper bounded by
    \[
    \lambda_n(L) < \sum_{i=1}^n \lambda_i(L) = \mathsf{Tr}(L)=\sum_{i=1}^n d_i = 2m,
    \]
    which implies that $\lambda_1(T_2)>1-2\alpha>-1$ as $\alpha\in(0,1)$ in our setting. Summarizing, we have
    \[
    -1<\lambda_1(T_2)\leq \cdots\leq \lambda_{n-1}(T_2) < \lambda_n(T_2) = 1 = \lVert T_2 \rVert.
    \]
    Since the operator norm is submultiplicative, we have
    \[
    \lVert \E[\A] \rVert \leq \lVert T_1 \rVert \times \lVert T_2 \rVert=1.
    \]
    However, if the upper bound 1 of $\lVert \E[\A] \rVert$ can be attained, then it is necessary that there exists a vector $v_\ast$ such that 
    \[v_{\ast} = v_n(T_1)=v_n(T_2).\]
    On the other hand, since $T_2$ commutes with $L$, it is well known that
    \[
    v_n(T_2) = v_1(L)=\frac{1}{\sqrt{n}}\1_n.
    \]
    However, it is readily observed that $\frac{1}{\sqrt{n}}\1_n^\T$ is not an eigenvector of $T_1$. We thus conclude that 
    \[
    \lVert \E[\A] \rVert < 1,
    \]
    completing the proof.
\end{proof}

\section{Derivation of Eq. (\ref{eq:y_infty_approx}) and discussions}\label{append:proof_y_infty}
First, let us use the regularization trick to introduce a regularized version of the Laplacian $L$ as
\[
\tilde L = L + \epsilon I_n,
\]
where $\epsilon$ is a small constant. Unlike the original $L$, such a regularized $\tilde L$ is invertible, but it shares the same set of eigenvectors of $L$. We also denote $\tilde T_2$ and $\E[\tilde \A]$ by replacing $L$ with $\tilde L$ in $T_2$ and $\E[\A]$, respectively. The introduction of the regularization makes the following analysis convenient (as matrices involved are invertible) and restores the original setting by letting $\epsilon\to 0$. It is straightforward to verify that this does not lead to any pathological behavior that would invalidate the approach. 

We then deal with the matrix inversion in Eq. (\ref{eq:y_infty_mat}). Note that 
\begin{align*}
    I_n - \E[\tilde A] & = \frac{\alpha}{m}\tilde L + \frac{\beta}{r}UU^\T\tilde T_2\\
    &= \frac{\alpha}{m}\tilde L\left(I_n +\frac{m\beta}{\alpha r} \tilde L^{-1}UU^\T\tilde T_2\right).
\end{align*}
By the Woodbury formula, we have
\begin{align*}
&\left(I_n +\frac{m\beta}{\alpha r} \tilde L^{-1}UU^\T\tilde T_2\right)^{-1} \\
= {}&I_n-\frac{m\beta}{\alpha r}\tilde L^{-1}U\left(I_r+\frac{m\beta}{\alpha r}U^\T\tilde T_2 \tilde L^{-1}U\right)^{-1}U^\T\tilde T_2.
\end{align*}
Therefore, $(I_n -\E[\tilde A])^{-1}U$ can be calculated as
\begin{widetext}
    \begin{align}\label{eq:mat_inv}
    (I_n -\E[\tilde A])^{-1}U 
    & =  \frac{m}{\alpha} \left(I_n-\frac{m\beta}{\alpha r}\tilde L^{-1}U\left(I_r+\frac{m\beta}{\alpha r}U^\T\tilde T_2 \tilde L^{-1}U\right)^{-1}U^\T\tilde T_2\right)\tilde L^{-1}U\notag\\
    &=\frac{m}{\alpha}\tilde L^{-1}U - \frac{m}{\alpha}\tilde L^{-1}U\left(I_r+\frac{m\beta}{\alpha r}U^\T\tilde T_2 \tilde L^{-1}U\right)^{-1}\left(\frac{m\beta}{\alpha r}U^\T\tilde T_2\tilde L^{-1}U\right)\notag\\
    &\overset{(\ast)}{=} \frac{m}{\alpha}\tilde L^{-1}U - \frac{m}{\alpha}\tilde L^{-1}U\left(I_r - \left(I_r+\frac{m\beta}{\alpha r}U^\T\tilde T_2\tilde L^{-1}U\right)^{-1}\right)\notag\\
    &= \frac{m}{\alpha}\tilde L^{-1}U\left(I_r+\frac{m\beta}{\alpha r}U^\T\tilde T_2\tilde L^{-1}U\right)^{-1},
\end{align}
where $(\ast)$ is due to the fact that $(I+M)^{-1}M=I-(I+M)^{-1}$, which can be directly verified by left multiplying $(I+M)$ to both sides.
\end{widetext}

To proceed, we use the spectral decomposition of $\tilde L$: 
\[
\tilde L =\sum_{i=1}^n \lambda_i(\tilde L)v_iv_i^\T,
\]
where $\{v_i\}$ is a set of orthonormal eigenvector of $L$ and
\[
\lambda_i(\tilde L) = \lambda_i + \epsilon
\]
with $\lambda_i$ being short for $\lambda_i(L)$, the $i$th smallest eigenvalue of $L$. Note that $\1_n=\sqrt{n}v_1$. We thus have
\begin{align*}
\1_n^\T \tilde L^{-1} = \sum_{i=1}^n\frac{1}{\lambda_i+\epsilon}\sqrt{n}v_1^\T v_iv_i^\T = \frac{\sqrt{n}}{\lambda_1+\epsilon}v_1^\T=\frac{\1_n^\T}{\epsilon},
\end{align*}
where we have used the fact that $v_1^\T v_i =\delta_{1i}$ and $\lambda_1=0$. Hence, 
\begin{align}\label{eq:1_r^T}
   \1_n^\T \tilde L^{-1}U =  \frac{\1_n^\T U}{\epsilon} = \frac{\1_r^\T}{\epsilon}.
\end{align}
Also, note that
\[
\tilde T_2\tilde L^{-1} = \left(I_n-\frac{\alpha}{m}\tilde L\right) \tilde L^{-1} = \tilde L^{-1} - \frac{\alpha}{m}I_n
\]
and $U^\T U=I_r$. It is straightforward to have
\begin{align}\label{eq:A2L_inv}
    U^\T \tilde T_2\tilde L^{-1} U
    =  U^\T \tilde L^{-1} U - \frac{\alpha}{m} I_r.
\end{align}
Combining Eqs. (\ref{eq:mat_inv}), (\ref{eq:1_r^T}), and (\ref{eq:A2L_inv}), we arrive at
\begin{align}\label{eq:T1_1}
    &\1_n^\T(I_n-\E[\tilde\A])^{-1}U \notag\\ 
    ={}& \frac{m}{\alpha\epsilon} \1_r^\T\left(I_r + \frac{m\beta}{\alpha r}\left(U^\T \tilde L^{-1} U - \frac{\alpha}{m} I_r\right)\right)^{-1} \notag\\
    ={}& \frac{m}{\alpha\epsilon} \1_r^\T\left(\left(1-\frac{\beta}{r}\right)I_r  + \frac{m\beta}{\alpha r}U^\T \tilde L^{-1} U\right)^{-1}\notag\\
    :={}&\frac{m}{\alpha\epsilon} \1_r^\T\tilde M^{-1},
\end{align}
where $\tilde M$ as defined above is clearly a full-rank $r\times r$ matrix. Applying the spectral decomposition of $\tilde L^{-1}$ again, we find that
\begin{align*}
\tilde M 
&= \left(1-\frac{\beta}{r}\right)I_r + \frac{m\beta}{\alpha r} U^\T \left(\sum_{i=1}^n \frac{v_i v_i^\T}{\lambda_i+\epsilon}\right) U\\
&= \frac{m\beta}{\alpha r \epsilon} U^\T v_1 v_1^\T U +  \left(1-\frac{\beta}{r}\right)I_r + \frac{m\beta}{\alpha r} U^\T \left(\sum_{i=2}^n \frac{v_i v_i^\T}{\lambda_i+\epsilon}\right) U\\
&:= \frac{m\beta}{\alpha n \epsilon} \frac{\1_r\1_r^\T}{r} + \tilde R.
\end{align*}
Note that all $\lambda_i>0$ for $i\neq 1$. Thus, the leading term of $\tilde M$ is $\frac{m\beta}{\alpha n \epsilon} \frac{\1_r\1_r^\T}{r}$ for small $\epsilon$.
Let $\{\hat v_1,\ldots,\hat v_r\}$ be a set of \emph{fixed} orthonormal basis vectors (independent of $\epsilon$) in $\mathbb{R}^r$ with $\hat v_1$ chosen as
\[
\hat v_1 = \frac{1}{\sqrt{r}}\1_r.
\]
Then we rewrite $\tilde M$ as
\[
\tilde M = \frac{m\beta}{\alpha n \epsilon} \hat v_1 \hat v_1^\T + \tilde R.
\]
Again, by the Woodbury formula, we have
\begin{align*}
    \tilde M^{-1} 
    = \tilde R^{-1} - \frac{\frac{m\beta}{\alpha n \epsilon} \tilde R^{-1}  \hat v_1 \hat v_1^\T \tilde R^{-1} }{1+ \frac{m\beta}{\alpha n \epsilon} \hat v_1^\T \tilde R^{-1} \hat v_1},
\end{align*}
which leads to 
\begin{align}\label{eq:T1_2}
    &\frac{m}{\alpha\epsilon} \1_r^\T\tilde M^{-1}  = \frac{m}{\alpha\epsilon} \1_r^\T\tilde R^{-1} - \frac{m}{\alpha\epsilon}\frac{ \1_r^\T\frac{m\beta}{\alpha n \epsilon} \tilde R^{-1}  \hat v_1 \hat v_1^\T \tilde R^{-1} }{1+ \frac{m\beta}{\alpha n \epsilon} \hat v_1^\T \tilde R^{-1} \hat v_1}\notag\\
    & = \frac{m}{\alpha\epsilon} \1_r^\T\tilde R^{-1} - \frac{m}{\alpha\epsilon}\frac{ \frac{m\beta}{\alpha r \epsilon} \sqrt{n}\hat v_1^\T \tilde R^{-1}  \hat v_1 \hat v_1^\T \tilde R^{-1} }{1+ \frac{m\beta}{\alpha n \epsilon} \hat v_1^\T \tilde R^{-1} \hat v_1}\notag\\
    & = \frac{m}{\alpha\epsilon} \1_r^\T\tilde R^{-1} - \frac{m}{\alpha\epsilon}\frac{ \left(\frac{m\beta}{\alpha n \epsilon} \hat v_1^\T \tilde R^{-1}  \hat v_1\right) \1_r^\T \tilde R^{-1} }{1+ \left(\frac{m\beta}{\alpha n \epsilon} \hat v_1^\T \tilde R^{-1} \hat v_1\right)}\notag\\
    & = \frac{m}{\alpha\epsilon} \1_r^\T\tilde R^{-1} - \frac{m}{\alpha\epsilon}\left(1 - \frac{1}{1+ \frac{m\beta}{\alpha n \epsilon} \hat v_1^\T \tilde R^{-1} \hat v_1}\right)\1_r^\T\tilde R^{-1}\notag\\
    &= \frac{\frac{m}{\alpha\epsilon} \1_r^\T\tilde R^{-1}}{1+ \frac{m\beta}{\alpha n \epsilon} \hat v_1^\T \tilde R^{-1} \hat v_1}.
\end{align}
By Eqs. (\ref{eq:T1_1}) and (\ref{eq:T1_2}), we have
\begin{align*}
    \1_n^\T (I_n- & \E[\tilde \A])^{-1} \E[\B]x_0\\
    ={}&\left(\1_n^\T(I_n-\E[\tilde \A])^{-1}U\right)\left(\frac{\beta}{r}U^\T x_0\right)\\
    ={}& \left(\frac{m}{\alpha\epsilon} \1_r^\T\tilde M^{-1}\right)\left(\frac{\beta}{r}U^\T x_0\right)\\
    ={}& \frac{\frac{m\beta}{\alpha\epsilon r} \1_r^\T\tilde R^{-1} U^\T x_0}{1+ \frac{m\beta}{\alpha n \epsilon} \hat v_1^\T \tilde R^{-1} \hat v_1}\\
    ={}& \frac{\frac{1}{\sqrt{r}}\hat v_1^\T \tilde R^{-1} U^\T x_0}{\frac{\alpha\epsilon}{m\beta}+\frac{1}{n}\hat v_1^\T \tilde R^{-1} \hat v_1}.
\end{align*}
Note that $U^\T x_0$ is a vector of dimension $r$. We use the basis $\{\hat v_k\}$ to expand it as
\[
U^\T x_0 = \sum_{k=1}^r c_k \hat v_k,
\]
where $c_k = \hat v_k^\T U^\T x_0$. In particular, 
\begin{align}\label{eq:c1}
c_1 = \hat v_1^\T U^\T x_0= \frac{\sum_{i=1}^r x_0(i)}{\sqrt{r}}.
\end{align}
We then have
\begin{align*}
   \1_n^\T (I_n-&\E[\tilde \A])^{-1}\E[\B]x_0\\
   = {}& \frac{\frac{1}{\sqrt{r}}\hat v_1^\T \tilde R^{-1} \sum_{k=1}^r c_k \hat v_k}{\frac{\alpha\epsilon}{m\beta}+\frac{1}{n}\hat v_1^\T \tilde R^{-1} \hat v_1} \\
   ={} & \frac{\frac{c_1}{\sqrt{r}}\hat v_1^\T \tilde R^{-1} \hat v_1 + \sum_{k=2}^r \frac{c_k}{\sqrt{r}}\hat v_1^\T \tilde R^{-1} \hat v_k}{\frac{\alpha\epsilon}{m\beta}+\frac{1}{n}\hat v_1^\T \tilde R^{-1} \hat v_1}.
\end{align*}
Inserting the expression of $c_1$ in Eq. (\ref{eq:c1}) and taking the limit $\epsilon\to 0$, we come to the conclusion that
\begin{align}\label{eq:y_infty}
    y_\infty&=\1_n^\T  (I_n-\E[\A])^{-1} \E[\B]x_0\notag\\ 
    &= \frac{nc_1}{\sqrt{r}} + \sum_{k=2}^r \frac{nc_k}{\sqrt{r}}\frac{\hat v_1^\T R^{-1} \hat v_k}{\hat v_1^\T R^{-1} \hat v_1}\notag\\
    & = \frac{n}{r}\sum_{i=1}^r x_0(i) + \sum_{k=2}^r \frac{nc_k}{\sqrt{r}}\frac{\hat v_1^\T R^{-1} \hat v_k}{\hat v_1^\T R^{-1} \hat v_1},
\end{align}
where
\[
R = \left(1-\frac{\beta}{r}\right)I_r + \frac{m\beta}{\alpha r} U^\T \left(\sum_{i=2}^n \frac{v_i v_i^\T}{\lambda_i}\right) U.
\]
Clearly, $R$ is positive definite. Hence, $R^{-1}$ exists and is also positive definite, and Eq. (\ref{eq:y_infty}) is thus always valid. Moreover, this result does not depend on the specific choice of $\hat v_2,\ldots, \hat v_r$ because
\[
\sum_{k=2}^r \hat v_k c_k=\sum_{r=2}^k \hat v_k \hat v_k^\T U^\T x_0 =(I_r - \hat v_1 \hat v_1^\T)U^\T x_0,
\]
which is fixed once $\hat v_1$ is selected. 

So far, there have been no approximations made, and the result is rigorous. On the right-hand side of Eq. (\ref{eq:y_infty}), the dependence on $\beta$ is via $R^{-1}$. We now investigate the situation when $\beta$ is small. We may rewrite $R$ as the sum of $I_r$ and an additional term proportional to $\beta$. That is, $R := I_r + \beta S$, where
\[
S = \frac{m}{\alpha r} U^\T \left(\sum_{i=2}^n \frac{v_i v_i^\T}{\lambda_i}\right) U - \frac{I_r}{r}. 
\]
Thus, for small $\beta$, we can view $R$ as a perturbation of the identity matrix $I_r$. Then $R^{-1}$ can also be seen as a perturbation of $I_r$, and can be written as
    \[
    R^{-1} = I_r - \beta S + o(\beta).
    \]
    To leading order in $\beta$, we have for $k\neq 1$ that 
    \begin{align*}
    \frac{\hat v_1^\T R^{-1} \hat v_k}{\hat v_1^\T R^{-1}\hat v_1} &\approx \frac{\hat v_1^\T I_r \hat v_k - \beta \hat v_1^\T S \hat v_k}{\hat v_1^\T I_r \hat v_1 - \beta \hat v_1^\T S \hat v_1}
    \approx -\beta \hat v_1^\T S \hat v_k\\
    &= -\beta \frac{m}{\alpha r} \hat v_1^\T U^\T\left(\sum_{i=2}^n \frac{v_iv_i^\T}{\lambda}\right) U\hat v_k,
    \end{align*}
    where we have used the fact that $\hat v_1^\T \frac{\beta}{r}I_r \hat v_k=0$ for $k\neq 1$.     Hence, for small $\beta$, we have
    \begin{widetext}
        \begin{align}\label{eq:y_infty_beta}
        y_\infty = \1_n^\T  (I_n-\E[\A])^{-1} \E[\B]x_0 = \frac{n}{r}\sum_{i=1}^r x_0(i) -\beta \frac{m}{\alpha r} \sum_{k=2}^r \frac{nc_k}{\sqrt{r}}\hat v_1^\T U^\T\left(\sum_{i=2}^n \frac{v_iv_i^\T}{\lambda_i}\right) U\hat v_k + o(\beta),
    \end{align}
    \end{widetext}
which is a more precise version of our first main result (\ref{eq:y_infty_approx}). Some discussions about the result are in order.

First, what do we mean by $\beta$ is small? Technically, we may require that the operator norm $\lVert \beta S \rVert\ll 1$. Note that
    \begin{align*}
    \lVert S \rVert &\leq \left\lVert \frac{I_r}{r} \right\rVert + \frac{m}{\alpha r}\left\lVert U^\T \left(\sum_{i=2}^n\frac{v_iv_i^\T}{\lambda_i}\right) U \right\rVert\\
    &\leq \frac{1}{r} + \frac{m}{\alpha r} \lVert U^\T \rVert \times \left\lVert \sum_{i=2}^n\frac{v_iv_i^\T}{\lambda_i} \right\rVert \times \lVert U\rVert\\
    &=\frac{1}{r} + \frac{m}{\alpha r \lambda_2},
    \end{align*}
    where the first step follows from the triangle inequality, and the second step is due to the submultiplicativity of the operator norm. Hence, we arrive at the sufficient condition (\ref{eq:small_beta}).

Second, there is an interesting special case where $x_0(i)$ is the same for $i=1,\ldots,r$, i.e., $U^\T x_0=c\1_r=c\sqrt{r}\hat v_1$ is a constant vector. In this case, $c_1=c\sqrt{r}$ and $c_k=0$ for $k\neq 1$, inserting which into Eq. (\ref{eq:y_infty}) yields
\begin{align*}
   &y_\infty = nc = \frac{n}{r}\sum_{k=1}^r x_0(i).
\end{align*}
This exact result shows that if all $r$ nodes in the subset have the same initial state, then they are able to drive the network to the desired state on average in the long-time limit, irrespective of the value of $\beta$.

\section{Analysis of $\Delta_t$}\label{append:convergence_rate}
We are interested in the regime where $t\gg 1$ and $\beta\ll 1$. As defined in the main text, $\Delta_t=\Delta_{t1}+\Delta_{t2}$. We first consider $\Delta_{t1}$. 

In general, $\E[\A]$ does not have a spectral decomposition. However, as defined above, $\E[\A]=T_1 T_2 = T_1^{1/2} (T_1^{1/2}T_2 T_1^{1/2}) T_1^{-1/2}$, where $T_1=I_n-\beta UU^\T/r$ and $T_2=I_n - \alpha L/m$. Hence $\E[\A]$ is similar to $T_1^{1/2}T_2T_1^{1/2}$, and 
\begin{align*}
\lVert \E[\A]\rVert &\leq \lVert T_1^{1/2}\rVert\times \lVert T_1^{1/2}T_2 T_1^{1/2} \rVert \times \lVert T_1^{-1/2}\rVert \\
&= 1 \times \lVert T_1^{1/2}T_2 T_1^{1/2} \rVert \times \sqrt{\frac{1}{1-\frac{\beta}{r}}}\\
&\approx \left(1+\frac{\beta}{2r}\right) \times \lVert T_1^{1/2}T_2 T_1^{1/2} \rVert.
\end{align*}
By similar arguments as in Appendix \ref{append:spectra}, one can show that the operator norm $\lVert T_1^{1/2}T_2 T_1^{1/2} \rVert<1$. Hence, if we approximate $\E[\A]$ by $T_1^{1/2}T_2 T_1^{1/2}$, then the relative error is on the order of $O(\beta)$. Moreover, since
\[
(T_1 T_2)^t = T_1^{1/2}(T_1^{1/2}T_2T_1^{1/2})^tT_1^{-1/2},
\]
we similarly have that the relative error incurred by using $(T_1^{1/2}T_2T_1^{1/2})^t$ to approximate $\E[\A]^t$ is also $O(\beta)$.  

As a real symmetric matrix, $(T_1^{1/2}T_2T_1^{1/2})^t$ can be written as
\[
(T_1^{1/2}T_2 T_1^{1/2})^t = \sum_{i=1}^n \mu_i^t u_i u_i^\T,
\]
where $\{\mu_i\}_{i\in[n]}$ are the eigenvalues of $T_1^{1/2}T_2 T_1^{1/2}$ and $\{u_i\}_{i\in[n]}$ are the corresponding orthonormal eigenvectors. The mode associated with the largest $|\mu_i|$ dominates for large $t$. We assume the truncation error (which depends on the spectrum of $L$) at a finite but large enough $t$ is negligible and do not contain it in the following discussion.  

Since we are interested in the case of a small $\beta$, $T_1^{1/2}=I_n+\left(\sqrt{1-\frac{\beta}{r}}-1\right)UU^\T$ can be seen as a perturbation of $I_n$, and $T_1^{1/2}T_2T_1^{1/2}$ can be seen as a perturbation of $T_2$:
\begin{align*}
T_1^{1/2}T_2T_1^{1/2} &\approx T_2 + \left(\sqrt{1-\frac{\beta}{r}}-1\right)\left( UU^\T T_2 + T_2 UU^\T \right)\\
&\approx T_2 - \frac{\beta}{2r} (UU^\T T_2 + T_2 UU^T),
\end{align*}
where we have omitted higher-order terms in $\beta$. Let $\{\lambda_i\}_{i\in[n]}$ be the set of eigenvalues of the Laplacian matrix $L$ and $\{v_i\}_{i\in[n]}$ be the set of corresponding eigenvectors. Then we can approximate the eigenvalues $\{\mu_{i}\}_{i\in [n]}$ as

\begin{align*}
\mu_i ={}& \left(1 - \frac{\alpha \lambda_i}{m}\right)- \frac{\beta}{2r}v_i^\T UU^\T\left(I_n - \frac{\alpha}{m}L\right)v_i\\
& -\frac{\beta}{2r}v_i^\T \left(I_n - \frac{\alpha}{m}L\right) UU^\T v_i + o(\beta).
\end{align*}
Notice that $\lambda_1=0$ and $v_1 = \frac{\1_n}{\sqrt{n}}$. We expect that when $\beta$ is sufficiently small, the largest $|\mu_i|$ is equal to 
\begin{align*}
\mu_n \approx  1 - \frac{\beta}{nr}\1_n^\T UU^\T \1_n = 1 - \frac{\beta}{n},
\end{align*}
where we have used the fact that $L\1_n = 0_{n\times 1}$. Similarly, the corresponding eigenvector is approximated by $v_1$ up to a term $\sim O(\beta)$. Hence, we conclude that, to leading order in $\beta$,
\begin{align*}
  \Delta_{t1} &\approx \left(1-\frac{\beta }{n}\right)^t \1_n^\T v_1 v_1^\T x_0 = \left(1-\frac{\beta }{n}\right)^t \1_n^\T x_0.
\end{align*}
This approximation of $\Delta_{t1}$ behaves well when the spectral gap of $L$ is not too small, i.e., when $\lambda_2$ is not too close to 0. Now, let us consider $\Delta_{t2}$. Note that 
\[
\E[A]^{t-1}+\cdots+I_n = (I_n-\E[A])^{-1}(I_n -\E[A]^t).
\]
Following the similar analysis, we also approximate $\Delta_{t2}$ for large $t$ and small $\beta$ as
\begin{align*}
    \Delta_{t2} &= -\1_n^\T (I_n-\E[\A])^{-1}\E[\A]^t \E[\B]x_0\\
    &\approx-\1_n^\T \left( \sum_{i=1}^n \frac{\mu_i^t}{1-\mu_i}u_iu_i^\T\right) \E[\B]x_0\\
    &\approx -\1_n^\T\frac{\mu_n^t}{1-\mu_n}u_nu_n^\T \frac{\beta}{r} UU^\T x_0\\
    &\approx -\frac{\beta}{r}\frac{(1-\beta/n)^t}{\beta/n}\1_n^\T UU^\T x_0\\
    & = -\left(1-\frac{\beta}{n}\right)^t\left(\frac{n}{r}\sum_{k=1}^r x_0(k)\right).
\end{align*}
Combining the results for $\Delta_{t1}$ and $\Delta_{t2}$, we have that
\begin{align}\label{eq:Delta_t}
\Delta_t &\approx \left(1-\frac{\beta}{n}\right)^t \left(\sum_{i=1}^n x_0(i)- \frac{n}{r}\sum_{k=1}^r x_0(k)\right)\notag\\
&\approx \left(1-\frac{\beta}{n}\right)^t\Delta_0.
\end{align}
We expect this approximation to be reasonably accurate for fairly large $\lambda_2$ and when $t\gg1$ and $\beta\ll1$. For future reference, we note that
\begin{align}\label{eq:approx_yt}
    &\1_n^\T\E[x_t] = \Delta_t + y_\infty\notag\\
    \approx{}& \left(1-\frac{\beta}{n}\right)^t \sum_{i=1}^n x_0(i) + \left(1 - \left(1-\frac{\beta}{n}\right)^t \right) \frac{n}{r} \sum_{k=1}^r x_0(k)\notag\\
    ={}& \left(1-\frac{\beta}{n}\right)^t \1_n^\T x_0 + \left(1 - \left(1-\frac{\beta}{n}\right)^t \right) \frac{n}{\beta}\1_n^\T \frac{\beta}{r} UU^\T x_0\notag\\
    ={}&\left(1-\frac{\beta}{n}\right)^t \1_n^\T x_0 + \left(1 - \left(1-\frac{\beta}{n}\right)^t \right) \frac{n}{\beta}\1_n^\T \E[\B] x_0\notag\\
    ={}& \left(1-\frac{\beta}{n}\right)^t\1_n^\T x_0 + \sum_{\tau=0}^{t-1}\left(1-\frac{\beta}{n}\right)^{t-1-\tau} \1_n^\T \E[\B] x_0.
\end{align}

\section{Probabilistic bounds for $Y_t$}\label{append:confidence_interval}
In this appendix, we aim to bound the (absolute) difference $|Y_t-y_t|=|\1_n^\T x_t -\1_n^\T \E[x_t]|$ with high probability, in an informative way. On a high level, we aim to show that $Y_t$ is a sub-Gaussian random variable and then provide an (approximate) sub-Gaussian parameter associated with it. Such a parameter can be used to construct an interval that covers $y_t$ with high probability, and it is informative in the sense that it explicitly depends on $\beta$. To this end, we will argue that the resetting-swapping dynamics considered in this work can be approximated by an effective one when $t$ is large and $\beta$ is small, in which the randomness can be handled more easily than in the original problem. By doing so, we sacrifice a bit of mathematical rigor, but gain clearer physical intuition, which captures the main feature of the process in a typical setting.

Note that, in order to calculate $Y_t$, we encounter terms in the form of $\1_n^\T\A_{t-1}\ldots\A_\tau$. To ease the notation, we denote
\[
R_\tau = e_{k_\tau}e_{k_\tau}^\T\text{  and  } S_\tau=(e_{i_\tau}-e_{j_\tau})(e_{i_\tau}-e_{j_\tau})^\T.
\]
Since $\1_n^\T S_\tau v=0$ for a generic vector $v$, we have
\begin{align*}
\1_n^\T\A_{t-1}&=\1_n^\T(I_n-\beta R_{t-1}-\alpha S_{t-1}+\alpha\beta R_{t-1}S_{t-1})\\
&=\1_n^\T(I_n-\beta R_{t-1}+\alpha\beta R_{t-1}S_{t-1}).
\end{align*}
Recall that $\sqrt{n}v_1=\1_n$ and $\sum_{i=1}^n v_iv_i^\T=I_n$. Thus,
\begin{align*}
   \1_n^\T&(I_n-\beta R_{t-1}+\alpha\beta R_{t-1}S_{t-1})\left(\sum_{i=1}^n v_iv_i^\T\right)\\
   ={}&\sqrt{n}v_1^\T(I_n-\beta R_{t-1}+\alpha\beta R_{t-1}S_{t-1})v_1v_1^\T\\
   & + \sum_{i=2}^n \sqrt{n}v_1^\T(I_n-\beta R_{t-1}+\alpha\beta R_{t-1}S_{t-1})v_iv_i^\T.
\end{align*}
Again, note that $v^\T S_{t-1}v_1=0$ for a generic $v$ and that for each $i\neq 1$, we have
\begin{align*}
    |v_1^\T(I_n-&\beta R_{t-1}+\alpha\beta R_{t-1}S_{t-1})v_i| \\
    ={}&|\beta(-v_1^\T R_{t-1} v_i + \alpha v_1^\T R_{t-1}S_{t-1}v_i)|\\
    \leq {}&  \frac{\beta(1+\sqrt{2}\alpha)}{\sqrt{n}}.
\end{align*}
Hence, for a unit vector $v$, we obtain
\begin{align*}
    &|\1_n^\T\A_{t-1}v - \left(v_1^\T(I_n-\beta R_{t-1})v_1\right) \1_n^\T v| \\
    \leq{}&\sum_{i=2}^n \sqrt{n}\frac{\beta(1+\sqrt{2}\alpha)}{\sqrt{n}}|v_i^\T v|\leq\beta(1+\sqrt{2}\alpha)\sqrt{n-1}\\
    ={}& O(\beta\sqrt{n}).
\end{align*}
It is sufficient that, if $\beta \sqrt{n}\ll 1$, then one can approximate $\1_n^\T\A_{t-1}$ by $\left[v_1^\T(I_n-\beta R_{t-1})v_1\right] \1_n^\T$ reasonably well, and
\begin{align*}
    v_1^\T(I_n-\beta R_{t-1})v_1 & = v_1^\T (I_n - \beta e_{k_{t-1}}e_{k_{t-1}}^\T)v_1 = 1-\frac{\beta}{n},
\end{align*}
which is deterministic and does not depend on the specific time $t-1$. However, we do not only want to approximate $\1_n^\T\A_{t-1}$ by keeping only the term associated with $v_1$; we also expect a similar approximation to be valid for $\1_n^\T\A_{t-1}\ldots\A_\tau$. That is,
\[
\1_n^\T\A_{t-1}\ldots\A_\tau \approx \left(1-\frac{\beta}{n}\right)^{t-\tau}\1_n^\T.
\]
A more precise error analysis is awaited, but the above approximation captures the ``first order" behavior. Based on this idea, the calculation for an approximate $\1_n^\T x_t$ is straightforward based on an effective dynamics:
\begin{align*}
    \1_n^\T x_t & \approx \left(1-\frac{\beta}{n}\right)^t\1_n^\T x_0 + \sum_{\tau=0}^{t-1} \left(1-\frac{\beta}{n}\right)^{t-1-\tau}\1_n^\T \B_\tau x_0.
\end{align*}
Thus, by Eq. (\ref{eq:approx_yt}), the fluctuation of $\1_n^\T x_t$ around its mean $\1_n^\T\E[x_t]$ is approximately
\begin{align*}
    \1_n^\T x_t -\1_n^\T \E[x_t]
\approx{}&\sum_{\tau=0}^{t-1} \underbrace{\left(1-\frac{\beta}{n}\right)^{t-1-\tau}}_{:=\omega(\tau+1)}\underbrace{\1_n^\T (\B_\tau - \E[\B])x_0}_{:=b(\tau+1)}\\
:={}& \omega^\T b,
\end{align*}
where $\omega$ and $b$ are two vectors of dimension $t$. Note that in this approximate expression for $Y_t-y_t$, the randomness only originates from $\B_\tau$. This is equivalent to an effective dynamics, which can be seen as a weighted sum of the resetting dynamics at each time step, and the weights decay exponentially with the temporal distance of each resetting event from the present.  

In the following, we will establish that $\omega^\T b$ is a sub-Gaussian random variable. Recall that, by definition, a sub-Gaussian random variable $X$ satisfies \cite{van2014probability} 
\[
\E[e^{s(X-\E[X])}]\leq e^{s^2\sigma^2/2}\text{ for all }s\in\mathbb{R},
\]
where $\sigma^2$ is a constant. We might denote $X\sim\subG(\sigma^2)$. It is well known that all bounded random variables are sub-Gaussian. In particular, if $X\in[a,b]$ almost surely, then
\[
X\sim\subG\left(\left(\frac{b-a}{2}\right)^2\right).
\]
A sub-Gaussian random variable $X\sim\subG(\sigma^2)$ is typically not too far away from its mean $\E[X]$, as quantified by the following concentration inequality \cite{van2014probability}:
\[
\p\left(|X-\E[X]|>\varepsilon\right)\leq 2\exp\left(-\frac{\varepsilon^2}{2\sigma^2}\right)
\]
for all $\varepsilon>0$. For each $i\in[t]$ and $\tau=i-1$, one can directly verify that
\begin{align*}
    b(i)&= \1_n^\T\left(\beta e_{k_\tau}e_{k_\tau}^\T-\frac{\beta}{r}\sum_{k=1}^r e_ke_k^\T\right)x_0 \\
    &= \beta x_0(k_\tau) - \beta\left(\frac{\sum_{k=1}^r x_0(k)}{r}\right)\\
    &\leq \beta\left(\max_{k\in[r]} x_0(k) - \frac{\sum_{k=1}^r x_0(k)}{r}\right).
\end{align*}
Similarly, we also have
\[
b(i)\geq \beta\left(\min_{k\in[r]} x_0(k) - \frac{\sum_{k=1}^r x_0(k)}{r}\right).
\]
Hence, for all $i\in[t]$, $b(i)$ is bounded and thus sub-Gaussian: 
\[
b(i)\sim\subG(\beta^2\R^2),
\]
where
\[
\R = \frac{1}{2}\left(\max_{k\in[r]}x_0(k) - \min_{k\in[r]}x_0(k)\right).
\]
Note that $b(i)$'s are independent. So $\omega^\T b$ is the weighted sum of $t$ independent sub-Gaussian random variables, and it is also sub-Gaussian with
\[
\omega^\T b \sim\subG(\lVert \omega\rVert^2 \beta^2\R^2).
\]
This can be verified by noticing that, for all $s\in\mathbb{R}$,
\begin{align*}
    &\E[e^{s(\omega^\T b-\omega^\T \E[b])}] = \E[e^{s\omega^\T b}]=\prod_{i=1}^t\E[e^{s\omega_ib_i}]\\
    \leq{}& \prod_{i=1}^te^{s^2\omega_i^2\beta^2\R^2/2}= e^{s^2\lVert \omega\rVert^2\beta^2\R^2/2}.
\end{align*}
Applying the tail probability bound for sub-Gaussian random variables \cite{van2014probability}, we have
\begin{align*}
    \p\left(|\1_n^\T x_t -\1_n^\T \E[x_t]|>\varepsilon\right) &\leq 2\exp\left(-\frac{\varepsilon^2}{2\lVert \omega\rVert^2\beta^2\R^2}\right)
\end{align*}
for all $\varepsilon>0$. Finally, note that
\begin{align*}
    \lVert \omega\rVert^2 &= \sum_{\tau=0}^{t-1} \left(\left(1-\frac{\beta}{n}\right)^{t-1-\tau}\right)^2=\sum_{\tau=0}^{t-1} \left(1-\frac{\beta}{n}\right)^{2\tau}\\
    &=\frac{1-(1-\beta/n)^{2t}}{1-(1-\beta/n)^2}.
\end{align*}
Since we are interested in the large-$t$ and then small-$\beta$ regime, we have
\[
\lVert \omega\rVert^2 \approx \frac{n}{2\beta}.
\]
In this case, the tail probability is approximately bounded by
\begin{align*}
    \p\left(|\1_n^\T x_t -\1_n^\T \E[x_t]|>\varepsilon\right) &\lesssim 2\exp\left(-\frac{\varepsilon^2}{n\beta\R^2}\right).
\end{align*}
Let the right-hand side be equal to $\delta\in(0,1)$. Then the result can be equivalently reinterpreted: with probability roughly as high as $1-\delta$, we have
\begin{align*}
    |\1_n^\T x_t -\1_n^\T \E[x_t]|\leq\sqrt{n\beta\ln\left(\frac{2}{\delta}\right)}\R.
\end{align*}
Moreover, if we require that $|\1_n^\T x_t -\1_n^\T \E[x_t]|\leq \varepsilon$ with probability $1-\delta$, where both $\varepsilon>0$ and $\delta\in(0,1)$ are prescribed, then we need 
\begin{align*}
    \beta\leq \frac{\varepsilon^2}{n\ln\left(\frac{2}{\delta}\right)\R^2}.
\end{align*}
\end{appendix}

\bibliography{reference}

\end{document}